\documentclass[fleqn,usenatbib]{mnras}
\usepackage{savesym}
\usepackage{amsmath}
\savesymbol{iint}
\savesymbol{iiint}
\usepackage{txfonts}
\restoresymbol{TXF}{iint}
\restoresymbol{TXF}{iiint}

\usepackage[T1]{fontenc}
\usepackage{ae,aecompl}

\usepackage{graphicx}	

\newcommand{\eg}{e.g.,\ }

\newcommand{\Msun}{M_{\odot}}

\newcommand{\kms}{km~s$^{-1}$}

\newcommand{\OI}{\ion{O}{i}}

\newcommand{\NaI}{\ion{Na}{i}}
\newcommand{\MgII}{\ion{Mg}{ii}}

\newcommand{\SiII}{\ion{Si}{ii}}
\newcommand{\SiIII}{\ion{Si}{iii}}

\newcommand{\SII}{\ion{S}{ii}}
\newcommand{\CaII}{\ion{Ca}{ii}}
\newcommand{\TiII}{\ion{Ti}{ii}}
\newcommand{\TiIII}{\ion{Ti}{iii}}
\newcommand{\VII}{\ion{V}{ii}}
\newcommand{\CrII}{\ion{Cr}{ii}}

\newcommand{\FeII}{\ion{Fe}{ii}}
\newcommand{\FeIII}{Fe{\sc iii}}
\newcommand{\CoII}{Co{\sc ii}}

\newcommand{\NiII}{\ion{Ni}{ii}}

\newcommand{\Cofs}{$^{56}$Co}
\newcommand{\Nifs}{$^{56}$Ni}
\newcommand{\Mej}{$M_{\rm ej}$}
\newcommand{\KE}{$E_{\rm kin}$}
\newcommand{\vph}{$v_{ph}$}
\newcommand{\Dm}{$\Delta m_{15}(B)$}
\newcommand{\lam}{$\lambda$}
\newcommand{\ab}{$\sim$}

\title[SN\,2007on \& SN\,2011iv]{On the type Ia supernovae 2007on and 2011iv: Evidence for Chandrasekhar-mass explosions at the faint end of the luminosity-width relationship. }

\author[C. Ashall]{
\parbox{\textwidth}{
\raggedright
C. Ashall$^{1,2}$,\thanks{E-mail:chris.ashall24@gmail.com}
P. A. Mazzali$^{2,3}$,
M. D. Stritzinger$^{4}$, 
P. Hoeflich$^{1}$,
 C. R. Burns$^{5}$,
C. Gall $^{4,6}$, 
E. Y. Hsiao$^{1}$, 
M. M. Phillips$^{7}$,
N. Morrell$^{7}$,
R.~J.~Foley$^{8}$
}\vspace{0.4cm}\\
\parbox{\textwidth}{
$^{1}$Department of Physics, Florida State University, Tallahassee, FL 32306, USA\\
$^{2}$Astrophysics Research Institute, Liverpool John Moores University, IC2, Liverpool Science Park, 146 Brownlow Hill, \\  Liverpool L3 5RF, UK\\
$^{3}$Max-Planck-Institut f\"ur Astrophysik, Karl-Schwarzschild-Str. 1, D-85748 Garching, Germany\\
$^{4}$Department of Physics and Astronomy, Aarhus University, Ny Munkegade 120, DK-8000 Aarhus C, Denmark\\
$^{5}$Observatories of the Carnegie Institution for Science, 813 Santa Barbara St., Pasadena, CA 91101, USA \\
$^{6}$Dark Cosmology Centre, Niels Bohr Institute, University of Copenhagen, Juliane Maries Vej 30, 2100 Copenhagen \O, Denmark\\
$^{7}$Carnegie Observatories, Las Campanas Observatory, 601 Casilla, La Serena, Chile \\
$^{8}$Department of Astronomy and Astrophysics, University of California, Santa Cruz, CA 95064, USA\\
}
\vspace{-0.75cm}
}

\begin{document}

\date{***}

\pagerange{\pageref{firstpage}--\pageref{lastpage}} \pubyear{2018}

\maketitle

\label{firstpage}

\begin{abstract}
Radiative transfer models of two transitional type~Ia supernova (SNe Ia)  have been produced using the abundance stratification technique. 
These two objects --designated SN\,2007on and SN\,2011iv-- both exploded in the same galaxy, NGC\,1404, which allows for a direct comparison.
SN\,2007on synthesised 0.25\,$\Msun$ of \Nifs\ and was less luminous than SN\,2011iv, which produced 0.31\,$\Msun$  of \Nifs. 
 SN\,2007on  had a lower central density ($\rho_{c}$) and higher explosion energy (\KE\ $\sim 1.3\pm$0.3$\times10^{51}$erg) than SN\,2011iv, 
 and it produced less  nuclear statistical equilibrium (NSE) elements (0.06\,$\Msun$). 
Whereas, SN\,2011iv had a larger $\rho_{c}$, which increased the electron capture rate  in the lowest velocity regions, and produced  0.35\,$\Msun$ of stable NSE elements. 
SN\,2011iv  had an explosion energy of \KE\ $\sim 0.9 \pm$0.2$\times10^{51}$erg.
Both objects had an ejecta mass consistent with the Chandrasekhar mass (Ch-mass),
and their observational properties are well described by predictions from delayed-detonation explosion models. 
Within this framework, comparison to the sub-luminous SN\,1986G indicates  SN\,2011iv and SN\,1986G have different
 transition densities ($\rho_{tr}$) but similar $\rho_{c}$.
Whereas,  SN\,1986G and SN\,2007on had a similar $\rho_{tr}$ but different $\rho_{c}$.
 Finally, we examine the colour-stretch parameter $s_{BV}$ vs. $L_{max}$ relation and determine that the 
bulk of SNe~Ia (including the sub-luminous ones) are consistent with
 Ch-mass delayed-detonation explosions, where the main parameter driving the diversity is $\rho_{tr}$. We also find  $\rho_{c}$ to be driving
 the  second order scatter observed at the faint end of the luminosity-width relationship. 
\end{abstract}


\section{Introduction}

Type Ia supernovae (SNe~Ia) are thought to originate from   the thermonuclear disruption of a Carbon-Oxygen (C-O) white dwarf(s) (WD) in a binary system.
The two favoured progenitor channels are the single degenerate scenario (SDS) and 
the double degenerate scenario (DDS).
In the SDS a C-O WD accretes material from a non-electron degenerate companion star, 
 either a main sequence, Helium or Red Giant star  \citep{Whelan73,Livne90,Nomoto91,Nomoto97,Woosley94,livne95}.
In the DDS the binary system consists of two C-O WDs \citep{WDWD,Webbink84}. 
Recently another scenario has been proposed where the orbit of two WDs  are perturbed
by a tertiary or quaternary companion, resulting in the direct collision of the WDs \citep{Rosswog09,Raskin09,Kushnir13,Dong15,Fang17}.
\vspace{15mm}

There are several potential explosion mechanisms for SNe Ia. 
In the merger of two WDs the decreasing orbital radius reaches the point where the
binary system becomes unstable and the two WDs spiral in on a dynamical time scale \citep{Dan14,Dan15}, and subsequently, the explosion is triggered by heat released during the merging process.
Whereas in the collision scenario the binary system looses angular momentum due to the presence of a third star, until the binary orbit becomes so eccentric that a collision can occur, with some non-zero impact parameter. In this case the relative speed is close to the escape velocity and a detonation is unavoidable \citep{Kushnir13}.
 
 When a WD approaches the Chandrasekhar mass (Ch-mass), the explosion can
be triggered by compressional heat near the WD centre \citep{Diamond15}. 
This is usually thought to be in the SDS, although
 the accreted material in this scenario
may also originate from a tidally disrupted WD in a double degenerate system  with
accretion on a secular time scale \citep{Piersanti03}. 
Finally, explosions of a sub Ch-mass WD in the SDS can be triggered by detonating the surface He layer,
which drives a shock wave igniting the centre, producing a so-called double detonation explosion  \citep{Nomoto80,Nomoto84,Woosley94,hk96,livne95,Jiang17}. 
Current evidence seems to suggest that explosions near the Ch-mass seem to favour observations (see e.g., \citet{Mazzali07,Hoflich17}).

In Ch-mass explosions the nuclear flame front proceeds as either a deflagration (less than the local sound speed) or a detonation (faster than the local sound speed), 
but recent work has shown it is not possible to reach a progenitor configuration which can 
achieve a pure detonation by an accreting WD \citep{Hoflich02b,Zingale05}. Also  pure 
deflagration models do not produce enough \Nifs\ to power the brightest SNe Ia \citep{Reinecke02};
moreover these models fully mix the elemental abundances in the ejecta, which is in direct disagreement 
with observations of a chemically layered structure \citep{Gamezo03,Stehle05,Jordan12}.
Therefore, one of the most popular models is the delayed detonation (a deflagration with a transition to a detonation) explosion. In a delayed detonation  explosion an initial deflagration phase  lifts the WD in its potential and partially unbinds the star, after which the flame front drops to a transition density ($\rho_{tr}$) and a detonation is triggered.
For a normally bright SNe Ia, to first order, $\rho_{tr}$ determines the amount of \Nifs\ synthesised in the ejecta, as there is more effective burning during the detonation phase \citep{Khokhlov91,Hoflich02,Gamezo03}.

A variation on the  delayed detonation scenario is  the pulsation delayed detonation (PDD) explosion.
This is where  the WD stays bound after the deflagration phase, and undergoes 
a pulsation prior to the deflagration to detonation transition. 
The amount of burning in the deflagration phase determines if the WD is strongly or weakly bound; 
 more burning reduces the binding energy of the WD. PDDs can be produced with varying mass and amplitude pulsations, and can explain some of the diversity amongst SNe Ia   \citep{Hoflich96,Dessart14,Stritzinger15}.
 
Hydrodynamical  models of deflagration fronts indicate that they are multi-dimensional in nature,
and strongly mixed by Rayleigh-Taylor instabilities \citep{Zeldovich70,Khokhlov95,Gamezo03,Livne05,Ropke07,Seitenzahl13}. 
However, from observations there is strong evidence that 
there is an additional process which is needed to suppresses the mixing. 
 Observations include:  (i) direct imaging of the supernova remnant s-Andromeda that shows a large 
Ca-free core indicative of high-density burning and limited mixing \citep{Fesen07,Fesen16,Fesen17},
(ii) significantly degraded spectral fits by models that have an injection of radioactive material into the Si/S layers \citep{Hoflich02}, and finally, (iii) the precense of pot-bellied line profiles observed in spectra obtained 1-2\,years after the explosion thought to be formed from a significant amount of stable Fe-group isotopes located in the central region of the ejecta 
\citep{Maeda10a,Hoflich04,Motohara06,Maeda10b,Stritzinger15,Diamond15}.

The cause of this suppression is  currently unknown, but magnetic fields may have something to do with it  \citep{Hoflich04,Penney14,Remming14},
 and show promising results \citep{Hristov17},  as well as rapid rotation in the initial WD \citep{Uenishi03,Yoon05}.  

Furthermore, observational evidence strongly suggests that SNe Ia are spherical. 
This includes: the overall spherical density distributions as shown by low continuum polarization \citep{Maund10,Patat12} and spherical SNe Ia remnants exhibiting evidence of a chemical layered structure  \citep{Rest05,Fesen07}.  Spherical delayed detonation models such as those from \citet{Hoflich17} naturally suppress the mixing and does account for all of these observational traits. 

To first order SNe~Ia can be characterised by their  luminosity-decline rate  (\Dm, stretch) relation \citep{Phillips99},
and the luminosity-color relation \citep{Tripp98}.
The amount of \Nifs\ in the ejecta determines the peak luminosity of the light curve \citep{Arnett82}, and the link between \Nifs, luminosity, 
and opacity determines the light curve shape \citep{Mazzali01}.
Using these empirically derived relations to calibrate the observed luminosity of SNe~Ia allows them to serve as powerful 
extragalactic distance indicators used to study the expansion history of the Universe. 
For example, the comparison of a low redshift and high redshift SNe~Ia samples provided the first direct evidence 
of accelerated expansion of the Universe  \citep[e.g.,][]{perlmutter,riess98,Riess16}. 

There are many sub-types of SNe Ia, including  fast declining and sub-luminous (1991bg-like SNe Ia), (\Dm$>$1.8\,mag), as well as  transitional SNe Ia.
Transitional SNe Ia sit in an area of parameter space between normal and 1991bg-like SNe Ia,
  have a \Dm\ between 1.6-2.0\,mag, and are rare objects \citep{Ashall16a}. 
Some examples of transitional SNe Ia are SN~1986G  \citep{Phillips87,Ashall16a},
2003hv \citep{Leloudas09},
iPTF\,13ebh  \citep{Hsiao15} and SN\,2015bp \citep{Srivastav17}. 

Theoretically, it has been proposed that 1991bg-like SNe~Ia could come from  a different type of 
progenitor system than that of normal SNe Ia. Some of the suggested scenarios are 
 sub Ch-mass explosions \citep{Stritzinger06,Blondin17}, 
 mergers of 
two WDs \citep{Pakmor10,Mazzali12}, or  delayed detonation explosion of 
Ch-mass WDs \citep{Hoflich02}. 
Furthermore, it has recently been suggested by \citet{Blondin17} that SNe Ia with \Dm>1.4\,mag and by \citet{Goldstein18} that SNe~Ia with \Dm>1.55\,mag,  can only be produced by sub Ch-mass explosions. 
Therefore, understanding transitional SNe~Ia and their link between normal and 1991bg-like objects is critical,  as it will help us to determine whether or not sub-luminous SNe~Ia come from a distinct population.

Transitional SNe~Ia are rare and have not yet been studied in detail, but one transitional SN~Ia, (SN~1986G) has 
been shown to be consistent a delayed detonation Ch-mass explosion with a  high central density ($\rho_{c}$) \citep{Ashall16b}. 
However, this is only one example,
and the subclass of transitional SNe Ia have a  diverse set of properties. Three important questions that need to be answered 
are:  (1) do any transitional SNe~Ia begin to diverge from the Ch-mass? (2) what causes the diversity in fast-declining SNe Ia? and (3) do transitional SNe Ia provide a smooth link between the normal and sub-luminous SNe Ia populations?

Today errors in the estimated cosmological parameters are  limited not by sample sizes, but by  a number of  systematics,
 such as errors in photometric calibration and understanding  the SNe~Ia intrinsic colour and dust relations. 
 While there are many efforts to address these problems such as \citet{Rheault10}, \citet{Stubbs10},
 and \citet{Sasdelli16}, another more promising route is to fully understand the physics of
 the explosions. This was one of the main objectives of the  {\it Carnegie Supernova Project} (CSP), and has 
 produced some promising results \citep[see \eg][]{Folatelli12, Burns14,Gall17}. 
  To further improve our  understanding of the Universe, and 
 to be able to determine which of the competing cosmological models correctly describes 
 its expansion history, 
 we need to obtain errors on the distances to SNe~Ia at the $\sim$1\% level.
To do this, understanding  what causes the intrinsic variations between SNe~Ia is key. With this in mind, 
 we set out to understand the explosion physics of two transitional SNe~Ia (2007on and 2011iv), 
 as well as the overall properties of the transitional SNe Ia population. 
 
Comprehensive datasets of SNe~2007on and 2011iv have recently been presented by   \citet{Gall17}. 
Both of these objects were located in NGC\,1404, with 
SN~2007on exhibiting a decline rate of \Dm=1.96$\pm$0.01\,mag and SN~2011iv exhibiting decline-rate of \Dm=1.77$\pm$0.01\,mag.
Both objects have features consistent with normal SNe~Ia including: (i) no  \TiII\ feature at $\sim$4500~\AA\ at maximum
 (ii) they both peak in the near-infrared bands prior to the optical bands, and (iii) they both  exhibit a secondary maximum in the near-infrared bands.
They also show features indicative of sub-luminous SNe Ia including (i) a high \SiII\ line ($\sim$5970 to 6355~\AA)
ratio\footnote{The \SiII\ line ratio is an indirect measurement of temperature as the \SiII\ 6355~\AA\ features becomes
saturated, and as \SiIII\ recombines, it populated more excited levels (the 5970~\AA\ feature) of \SiII. This has the affect of increasing the 
line ratios \citep{Nugent95,Hachinger08}.} (ii) and a fast-evolving light curve.
These characteristics suggest that their photospheres are in the temperature regime located just above those of sub-luminous SNe~Ia.

The \Nifs\ mass inferred from the peak luminosity, using Arnett's rule, estimated from the UVOIR light curves presented by \citet{Gall17},  
correspond to 0.25$\pm$0.05$\,M_{\sun}$ for SN~2007on and 0.42$\pm$0.06\,$M_{\sun}$  for SN~2011iv.
This compares to 0.32\,$M_{\sun}$ for SN\,2007on and 0.37$\,M_{\sun}$ for SN\,2011iv from \citet{Hoflich17},
 and
 to 0.17\footnote{Calculated below 4000\,\kms\ using a 2 component model.}\,$M_{\sun}$ for
  SN~2007on and
 0.41\footnote{Calculated using a 1-zone model. }/0.32\footnote{Calculated using a 2 component model.}\,$M_{\sun}$  for SN~2011iv  as estimated from
  the nebular phase models of  \citet{Mazzali17}.

One intriguing characteristic found when comparing the $B-V$ colour evolution of  SN~2007on and SN~2011iv is the fact that at early 
times the latter is brighter and bluer than the former, but after +20~days relative to $B$-band maximum and extending past +85~days,  
SN 2007on is 0.12~mag bluer than SN 2011iv. 
 \citet{Gall17} speculate  this is due to differences in $\rho_{c}$ with their delayed detonation  models 
predicting bluer colours at late time as a function of decreasing $\rho_{c}$. 
This 0.12 mag difference in $B-V$ colours therefore suggests  that SN~2007on had a  $\rho_{c}$ about half of that of SN~2011iv.
Another intriguing feature about SN~2007on is its multiple emission peaks in the nebular phase \citep{Dong15}.
In a companion paper \citet{Mazzali17} demonstrate that these
peaks are possibly due to  two different components, one redshifted and one blue-shifted, with similar ejecta and  \Nifs\ masses but different degrees of ionisation. 
These components suggest that SN\,2007on  is either a very off-centre delayed detonation  explosion of a (near) Ch-mass WD, or   
a  collision of two similar mass WDs. 
In either case,  both scenarios agree that SN~2007on had a lower $\rho_{c}$ than SN~2011iv.

In this paper we perform a detailed radiative transfer synthetic spectral analysis on both SN~2007on and SN~2011iv, in order to explain their differences
and propose a progenitor scenario for their explosions. 
The method we use  is the abundance stratification technique. 
Abundance stratification utilises the fact that 
as time passes deeper and deeper layers of the expanding ejecta are revealed. Therefore a time series of spectra can 
be used to determine key properties of the SN~Ia, including the abundance distribution 
(in velocity and mass space) of various elements in the ejecta, which reveal themselves through spectral line features. This technique has been used for many SNe Ia, including SNe Ia
2002bo \citep{Stehle05}, 2003du \citep{Tanaka11}, 2004eo \citep{Mazzali08}, the very nearby SNe 2011fe \citep{Mazzali13} and
 2014J \citep{Ashall14}, and the transitional SN 1986G \citep{Ashall16b}, as well as core-collapse SNe \citep[\eg][]{Ashall17,Prentice17,Mazzali17}. It was also applied to SN\,2011ay \citep{Barna17} using a different moddeling code.
In this paper we  demonstrate how both SNe~2007on and 2011iv fit into the transitional and 
sub-luminous SNe Ia paradigm. Finally, we review the area of transitional and fast-declining SNe Ia, as well as the $s_{BV}$ vs. $L_{max}$ relation.

\section{modelling technique}
The abundance stratification technique makes use of the fact that a SN 
explosion can be split into two physically distinct parts: the photospheric and nebular phases. 
The results from the photospheric phase modelling can be used as an input for the nebular phase
modelling. This allows us to probe the full abundance distribution in velocity and mass space.

A SN ejecta can be thought to be in homologous expansion $\sim$10\,s after explosion. This is estimated by the equation
$r=v_{ph}\times t_{exp}$, where $r$ is the 
distance from the centre of the explosion, $v_{ph}$ is the photospheric velocity
and $t_{exp}$ is the time from explosion. Hence, as time passes deeper and deeper layers of the explosion can be seen.

At early times the ejecta are dense enough to emit continua within itself, producing an effective photosphere. 
This photosphere is actually a line-dominated pseudo photosphere, as line blanketing is the dominant source of opacity 
in a SN ejecta \citep{Pauldrach96}. Therefore, the Schuster-Schwarzchild approximation is used in our code.
 This approximation 
assumes that radiative energy is emitted from an inner blackbody. The radiation transport is then calculated above this
pseudo photosphere. The Schuster-Schwarzchild approximation is very useful as it does not require an in-depth knowledge 
of the radiation transport below the photosphere, but it still yields good results. However, at later times after bolometric maximum, 
this approximation can cause excess flux in the red ($>6500\AA$), but as most of the SNe Ia lines are in the blue this excess flux will not
affect the results with respect to abundances derived. 

The code is a 1D Monte carlo (MC) radiative transport code \citep{Mazzali93,Lucy1999,mazzali2000}. 
It simulates the emission of photon packets at a photosphere, and traces their evolution through a SN atmosphere. These packets can undergo Thomson scattering
and line absorption. If the latter occurs, the packet is immediately reemitted, following a branching scheme. 
This branching scheme allows fluorescence (blue to red) and reverse fluorescence (red to blue) to take place. 
Both of these are critical in forming a SN Ia spectrum. Fluorescence makes the optical part of the spectrum
and reverse fluorescence the UV.  Packets that  scatter back into the photosphere are reabsorbed,
therefore the blackbody temperature ($T_{BB}$) is iterated to match the input bolometric luminosity ($L_{bol}$)
given the back scattering rate.

 Ionisation and excitation are treated using a modified nebular approximation to account for non-local 
thermodynamic equilibrium (NLTE) effects caused by the radiation field \citep{Mazzali93, mazzali2000}. 
The radiation field and the state of the gas are iterated until convergence is reached. The final spectrum is 
computed using the formal integral to reduce Poisson noise \citep{Lucy1999}.

The aim of the abundance stratification technique is, starting with the earliest spectrum, to vary input parameters
such as abundance, $v_{ph}$ and $L_{bol}$ to produce optimally fitting synthetic spectra. 
Before the free parameters can be varied, the density profile, distance, and extinction to the SNe must be set. 
Previous work on spectral modelling of normal SNe Ia tended to use the W7 density profile \citep{Stehle05,Ashall14}. 
The W7 model is a fast deflagration which synthesised 0.5-0.6\,$\Msun$ \Nifs\ in the inner layers of the star, and has a  kinetic energy (\KE) of $1.3\times10^{51}\,$erg. 
However, in this work we start with  the same density profile required for SN 1986G, due to its similarities to SN 2007on and SN 2011iv.
SN 1986G has been shown to be consistent with a  low energy  W7-like density \citep{Ashall16b}.
 For convenience we will call this model W7e0.7 as it has a \KE\ 70\% of the standard W7 model, 
i.e., \KE\ =$0.9\times10^{51}$\,erg. When the W7e0.7 produces inadequate results 
(i.e., a poor fit to the observed spectra) we also test the 
W7 and W7e2 (\KE\ =$2.6\times10^{51}$\,erg) density profiles.
 This is done to demonstrate how different values of \KE\ will affect the models\footnote{We choose to stick to 3 models with a range of properties 
 as it gives us a range of results to choose from. It also avoids the subjective nature of
 determining the best fits, 
 which can occur when a fine grid is used.}.

\begin{figure}
\centering
\includegraphics[scale=0.8]{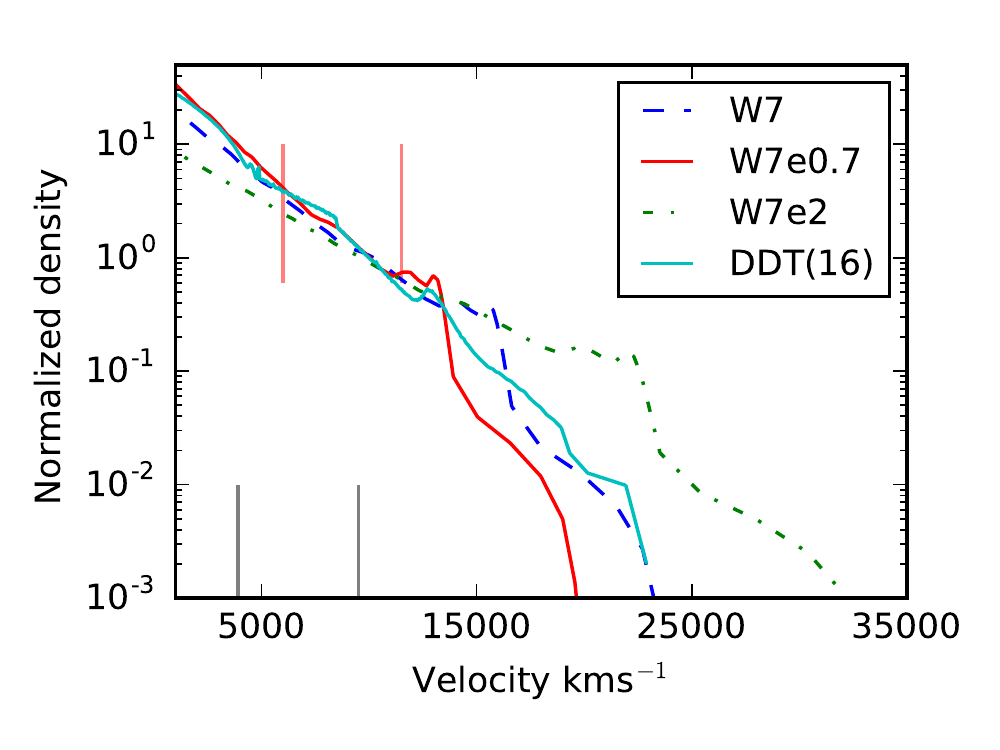}
\caption{The W7 (dashed blue),  W7e0.7 (solid red), W7e2 (dashed-dotted green), and DDT(16) density profile from \citet{Hoflich17},
normalized to 10000\kms. The vertical  lines show the $v_{ph}$ range
for SN\,2007on (back) and SN\,2011iv (red).}
\label{fig:densityprofile}
\end{figure}

 We note that the density profile of the deflagration model W7 is very similar to the structure of 
 a delayed detonation model for transitional SNe Ia. In both cases material with more than 14000\,\kms\ 
 is largely unburnt and as a consequence the density structure shows a  `bump' at the same velocity, 
 and higher densities compared to normal-bright delayed detonation models.  
  Figure~\ref{fig:densityprofile} shows the density profiles
 of the W7, W7e0.7, W7e2 models as well as DDT(16) from \citet{Hoflich17}. It should be noted how similar the
 W7e0.7 and DDT(16) density profiles are, especially in the spectral formation region. Therefore the assumption 
 of using a W7-like profile holds. 
 However, the similarity between W7-like profiles and delayed detonation models is only true for  `transitional' SNe Ia.

Transitional and sub-luminous SNe Ia are highly sensitive to physical second order parameters 
such as $\rho_{c}$ \citep{Hoflich17},
 therefore it is uncertain how accurately these objects
can be used as distance indicators.
It is known that sub-luminous and transitional SNe Ia tend to favour early type galaxies \citep{Hamuy96,Ashall16a}, with
SNe~2007on and 2011iv being no exception.  
Both were located in the same early type galaxy (NGC~1404), and suffer from 
almost no detectable host galaxy extinction\footnote{All of the spectra in this 
work were corrected for $E(B-V)_{MW}$=0.01\,mag \citep{Gall17}.}.
 This makes them an excellent case to test the accuracy  
of  current methods for calculating distances to transitional SNe Ia.
\citet{Gall17} found that even after correcting for 
stretch and colour (extinction) both objects exhibit significantly different peak magnitudes, 
and the relative distances derived to them vary significantly.
The distances they derived to the SNe differ by 14\% if the $B$-band was used, and 
9\% if the $H$ band was used. If SNe Ia are to be utilised as more precise cosmological rulers,
understanding the diversity in their physics, and linking this back to light curve fitting tools, is essential. 
Utilising explosion models and physical second order effects, 
\citet{Hoflich17}  determined the distance modulus 
to the SNe to be consistent with each other and in the range of 31.2--31.4\,mag.

Here, we use the independent method of 
spectral modelling to refine the distance to the SNe. By producing one-zone models, 
the distance to the SNe can be varied within a range until an optimal fit for both objects is determined. 
For SN 2007on and SN 2011iv a distance modulus of 31.2$\pm$0.2\,mag was found to produce the best fit 
i.e.  the correct line ratios, ionisation and velocity (see Appendix \ref{app:dista}).
 This is fully consistent with the value calculated in 
\citet{Hoflich17}. Furthermore, for SN\,2011iv (the object with the more complete dataset) \citet{Hoflich17}
determine the same distance modulus derived in this work (i.e., 31.2\,mag). 

Once the extinction, distance and density profile are set the main part of the modelling procedure can take place. 
Starting with the earliest spectra, $L_{bol}$ and $v_{ph}$ are first determined  and a typical 
abundance distribution is assumed.  There are
usually 2 or 3 synthetic shells placed above the photosphere, in order to produce a stratified
abundance distribution at high velocities. The abundances of the photospheric shell
are varied until an optimal solution is found. 
The next spectrum is then modelled, and only the abundances inside 
the previous photosphere are changed. Abundances in the outer layers can affect the synthetic spectra at later epochs, as elements well above the photosphere can produce opacity
 (usually at a higher velocity relative to the observations)
if there is large enough masses and densities in the outer shells of models. 
If this is the case iteration is required on the abundances of the elements in the outer layers.
Abundances given in this work have an error of $\pm$25 per cent, and photospheric velocities have an error of $\pm$15 per cent.
A full detailed analysis of the errors from the abundance stratification method can be found in \citep{Mazzali08}.
We note that with this method the existence of one good fit does not exclude other scenarios.

 \section{Photospheric phase}
In this section we present the photospheric phase models of SNe~2007on and 2011iv. 
All of the data used in this paper were first published in \citet{Gall17}. A summary of the spectra, and its
phase relative to $B$-band maximum, can be found in Tables \ref{table:data11iv} $\&$ \ref{table:data07on}. 

 \subsection{SN\,2007on}

\begin{figure*}
\centering
\includegraphics[scale=0.9]{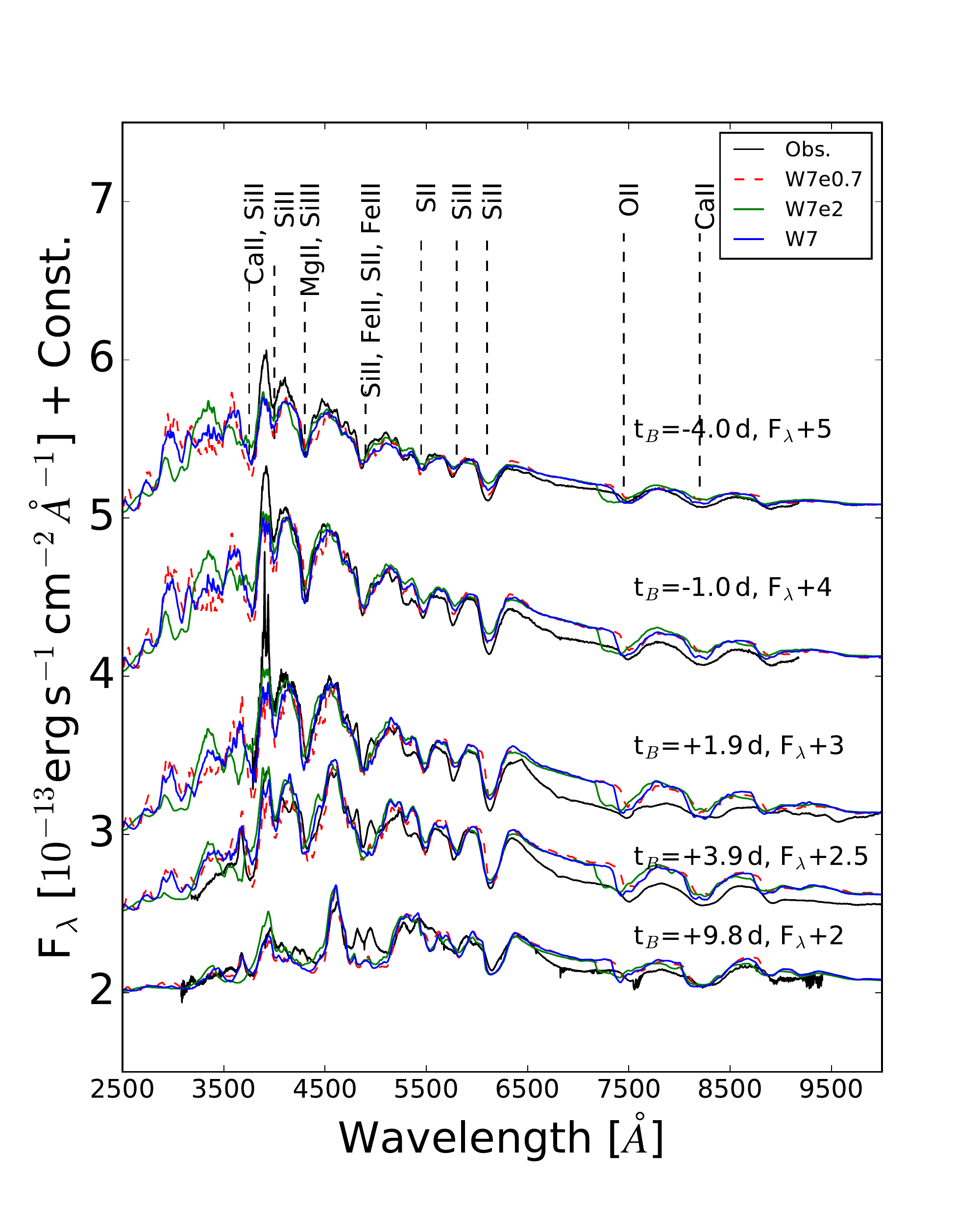}
\caption{Time-series of  photospheric phase visual-wavelength spectroscopy of SN\,2007on (black lines) 
compared with models produced using three density profiles W7 (blue solid line), W7e0.7  
(red dashed line) and W7e2 model (green solid line). The time relative to $B$-band maximum and the constant flux offset is plotted next to each spectrum.
}
\label{fig:07on}
\end{figure*}

Five photospheric phase spectra of SN~2007on, from
12.8 to 26.6\,d after explosion were modelled (see Figure  \ref{fig:07on}).  

The spectra are dominated by the standard SN\, Ia features, including  (\CaII\ $\lambda\lambda$3933,\,3968,
 \SiII\ $\lambda\lambda$4128,\,4130, \MgII\ $\lambda$4481,
\SiIII\ $\lambda\lambda$4552,\,5978,\,6347,\,6371, \FeII\ $\lambda\lambda$4923,\,5169, \FeIII\  $\lambda$5156,
 \SII\ $\lambda$$\lambda$5432,\,5453,\,5606,
  \OI\ $\lambda\lambda$7777,\,7774,\,7774, and  \CaII\ $\lambda\lambda$8498,\,8542,\,8662). 
The input parameters of the models can be found in Table \ref{table:07oninput}. 
The photospheric velocity of SN\,2007on covers a range from 3900--9500\,\kms, and
a rise time of 16.8$\pm$0.5\,d has been used, implying that SN~2007on exploded on JD 2,454,403.5. 

Due to the lack of early and UV data of SN~2007on we cannot test the metalicity of the progenitor, or the composition of the 
 outermost layers of the ejecta, 
but we can determine the basic properties and abundances of the explosion. 
We note that although the inner layers of SN~2007on are made of two components (see \citet{Mazzali17}),
 at early times these components
 will be hidden below the optically thick 
photosphere, hence our modeling results are still valid.  

The fits from the model made with the W7e0.7 density profiles (see Figure \ref{fig:07on}) are reasonable,
but do show signs of being too hot: 
the spectrum has a large \SiIII\ absorption at \ab4400\AA. Therefore, we made models with the W7 and W7e2 density profiles. 
The W7e2 model has too much oxygen absorption at high velocities, as seen in the \ab7500\AA\ feature. 
This is caused by an enhanced density in the outermost layers  where oxygen is dominant. 
The W7 model produces the best fits, and is therefore our preferred choice. Below we discuss the results from the W7 model at
three distinct epochs.

\subsubsection{$-$4.0\,d}
The earliest spectrum of SN~2007on was obtained 12.8\,d after explosion. The $v_{ph}$ at this epoch is 9500\,\kms, at this 
velocity there is $\sim$0.6\,\Mej\ of material above the photosphere. 
Going from blue to red, the main lines that contribute to the spectrum  are
\CaII\ \lam \lam3934,\,3968 \SiII\ \lam \lam4131,\,4128, \MgII\ \lam \lam4481.13,\,4481.32, \SiIII\ \lam \lam4553,\,4568,\,4575,
\SiII\ \lam \lam5041,\,5056, \FeII\ \lam \lam5018,\,5169, \SII\ \lam \lam5009,\,5032,\,5212,\,5432,\,5453,\,5606,
\SiII\  \lam \lam5957,\,5978,\,6347,\,6371, \OI\ \lam \lam7772,\,7774,\,7775, \CaII\ \lam \lam8542,\,8662 and \MgII\ \lam \lam9218,\,9244.
Even at this early epoch there is very little \FeIII\ absorption in the model, as is seen in the narrowness of the 4800\AA\ feature (see Figure  \ref{fig:07on}).

\subsubsection{+1.9\,d}
The model at 18.7\,d after explosion has a $v_{ph}$ of 8700\,\kms. The spectrum is similar to the one obtained at 
$-$4.0\,d, 
except for a  few notable differences, which are i) the feature at 4400\AA\ contains no contribution from \SiIII\ \lam \lam4553,\,4568,\,4575,
ii) \CoII\ \lam\lam4136,\,4660  lines are seen in the 4400\AA\ feature, this \CoII\ is produced from the decay of \Nifs\ located above
 the photosphere, iii) there are blends of \CoII\ \eg \lam\lam3446,\,3501,\,3621 lines in the near-UV.

\subsubsection{+9.8\,d}
 The model at 26.6\,d after explosion shows a lot of evolution with respect to the previous epoch, especially in the blue part of the spectrum. 
 Due to the adiabatic expansion of the ejecta the spectrum is significantly redder and cooler than previous epochs. 
 Most of the Ti abundance at this phase is singly ionised (i.e. \TiII\ \lam \lam3446,\,3501,\,3621), whereas in the model at +1.9\,d almost 
 all of the Ti was  doubly ionised (i.e. \TiIII). 
The 3000$-$4000\AA\ region contains  blends of 
 \TiII\ and \CoII\ (\eg \lam \lam3446,\,3501,\,3621) 
 lines as well as \CaII\  \lam \lam3934,\,3968. 
The strongest lines in the feature at \ab4300\AA\  are \SiII\ \lam \lam4131,\,4128 and  \TiII\  \lam \lam4395,\,4444,\,4469,
however, this feature also contains blends of  lines from many ions including \CoII, \NiII, \MgII, \CrII, and \FeII. 
 The feature at 4900\AA\ contains significant contribution from \FeII\ \lam \lam4924,\,5018,\,5169, \SiII\ \lam \lam5041,\,5056 
and \FeIII\ \lam \lam5156,\,5127,\,5074.

\begin{table}
 \centering
 \caption{Input parameters for the photospheric models of SN 2007on.}
   \begin{minipage}{80mm}
  \begin{tabular}{cccccc}
  \hline
   Epoch&phase&velocity& UVOIR L$_{\rm max}$\\
  $t_{rise}$\footnote{Days since inferred explosion time (JD 2,454,403.5).} &$t_{peak}$\footnote{Rest frame time relative to $B$-band maximum (JD 2,454,420.4).} &\vph&log$_{10}$(L$_{max}$)&\\
    \hline
  days&days&km\,s$^{-1}$&erg s$^{-1}$\\
  \hline
 12.8&$-$4.0&9500&42.37\\
 15.8&$-$1.0&9200&42.50\\
 18.7&$+$1.9&8700&42.50\\
 20.7&$+$3.9&6600&42.44\\
 26.6&$+$9.8&3900&42.19\\
\hline
\end{tabular}
\label{table:07oninput}
\end{minipage}
\end{table}

 \subsection{SN\,2011iv}
\begin{figure*}
\centering
\includegraphics[scale=0.4]{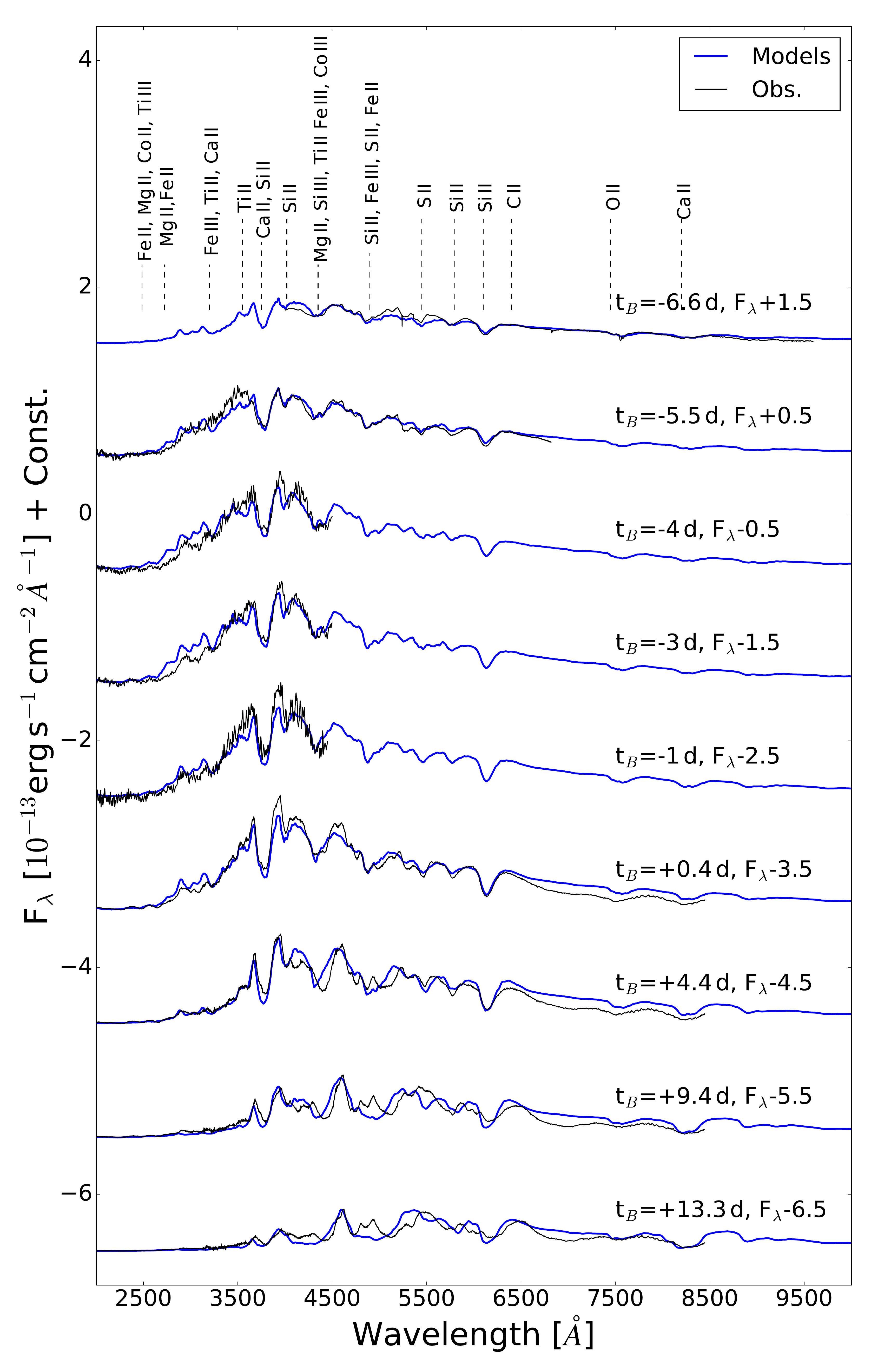}
\caption{Time-series of  photospheric phase visual-wavelength spectroscopy of SN\,2011iv (black lines) compared to the synthetic models (blue lines).  The models were computing  using the W7e0.7 density profile. The time relative to $B$-band maximum and the constant flux offset is plotted next to each spectrum.
 }
\label{fig:11ivphot}
\end{figure*}

Figure \ref{fig:11ivphot} contains the comparison between the observations  (black) 
and the photospheric phase models (blue) of SN\,2011iv, produced with the
 W7e0.7 density profile.  As these models produce good fits to the data, and \citet{Ashall16b}
 test multiple density profiles for transitional SNe Ia, we do not test other density profiles for
 SN\,2011iv. The spectra 
are dominated by normal SNe Ia features  (\CaII\ $\lambda\lambda$3933,\,3968, \SiII\ $\lambda\lambda$4128,\,4130, \MgII\ $\lambda$4481,
\SiIII\ $\lambda\lambda$4552,\,5978,\,6347,\,6371, \FeII\ $\lambda\lambda$4923,\,5169, \FeIII\  $\lambda$5156,
 \SII\ $\lambda$$\lambda$5432,\,5453,\,5606,
  \OI\ $\lambda\lambda$7777,\,7774,\,7774 and  \CaII\ $\lambda\lambda$8498,\,8542,\,8662), and were observed from $-$6.6\,d to +13.3\,d relative to $B$-band maximum light.
 
Generally the models fit the observed spectra well, including in the UV. The input for the models of SN 2011iv can be seen in 
Table \ref{table:11ivinput}. The photospheric velocity covers a range of 5500\,\kms, with the largest photospheric velocity 
being 11500\,\kms. In our models the bolometric luminosity peaks at 18.4\,d after explosion. At this point the photosphere is approximately
half way through the ejecta, and has a velocity of 9500\,\kms. 

A SN Ia spectrum develops over the course of months, and except at very early times, there is only a small evolution over the course
of a single day. Therefore, we discuss the models and spectra of SN 2011iv at four epochs, $-5.5$\,d, $+0.4$\,d, $+4.4$\,d, $+13.3$\,d.   

\begin{table}
 \centering
 \caption{Input parameters  for the photospheric models of SN\,2011iv.}
  \begin{minipage}{80mm}
  \center
  \begin{tabular}{ccccc}
  \hline
   Epoch&phase&velocity& UVOIR L$_{\rm max}$\\
  $t_{rise}$\footnote{Days since inferred explosion time (JD 2,455,924.6).} &$t_{peak}$\footnote{Rest frame time relative to $B$-band maximum (JD 2,455,906.1).}&$v_{ph}$&log$_{10}$(L$_{\rm max}$)\\
  \hline
  days&days&km\,s$^{-1}$&erg s$^{-1}$\\
  \hline
  11.8&$-$6.6&11500&42.64\\
  12.9&$-$5.5&10200&42.81\\
  14.4&$-$4.0&9700&42.87\\
  15.4&$-$3.0&9600&42.91\\
  17.4&$-$1.0&9550&42.93\\
  18.8&$+$0.4&9500&42.95\\
  22.8&$+$4.4&8200&42.90\\
  27.8&$+$9.4&6600&42.77\\
  31.5&$+$13.3&6000&42.65\\
\hline
\end{tabular}
\label{table:11ivinput}
\end{minipage}
\end{table}

\subsubsection{$-$5.5\,d}
\begin{figure*}
$\begin{array}{cc}
\includegraphics[scale=0.7]{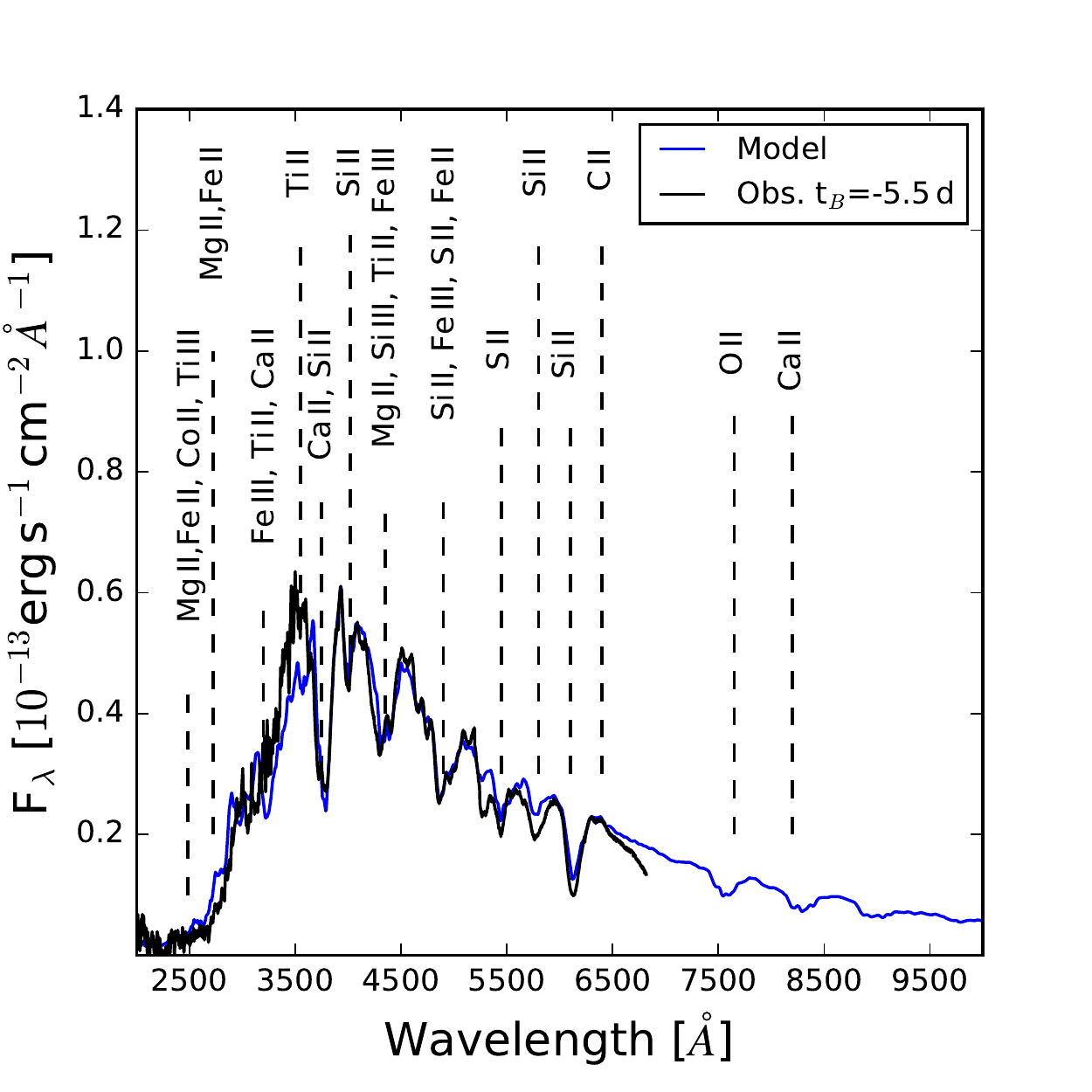} & \includegraphics[scale=0.7]{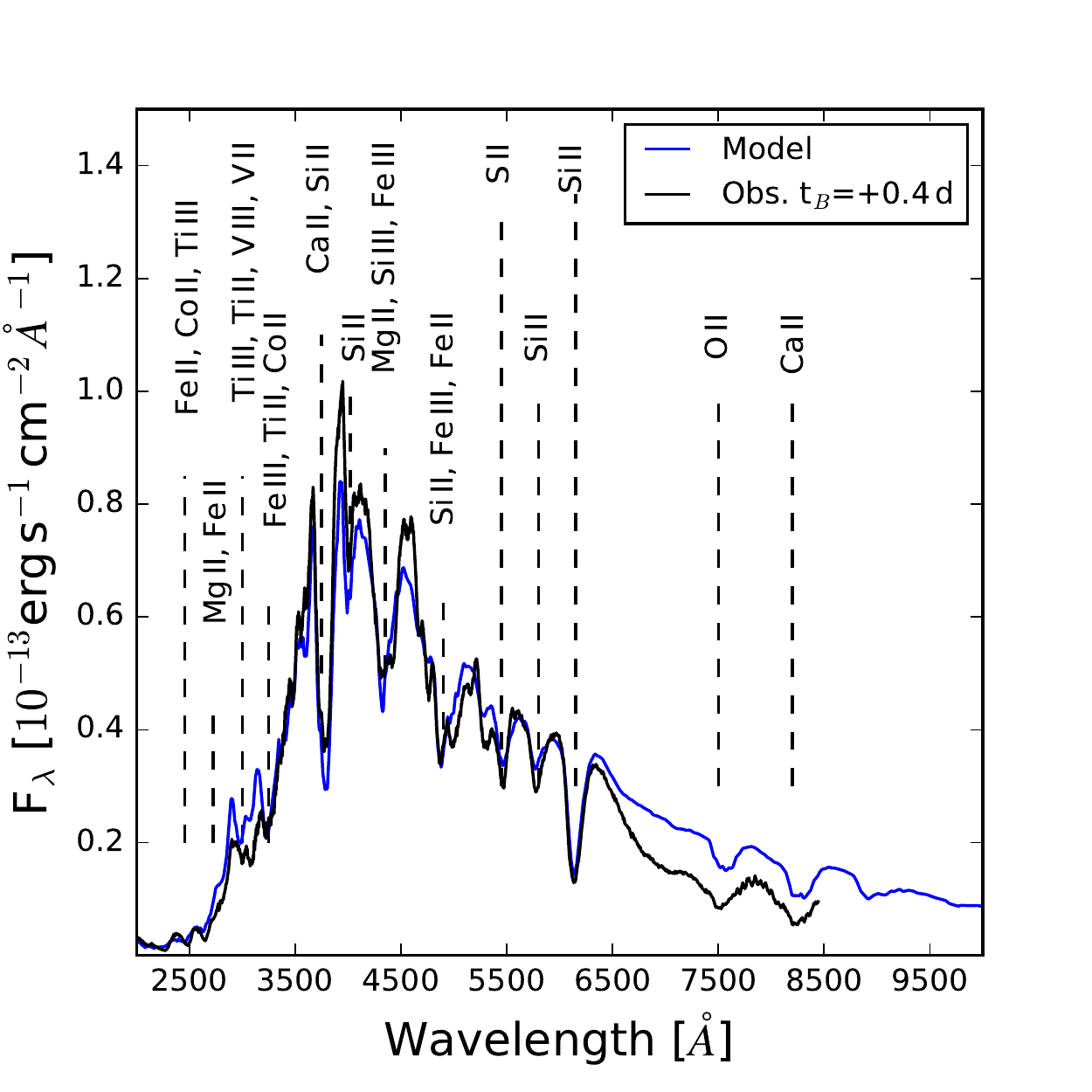} \\
\includegraphics[scale=0.7]{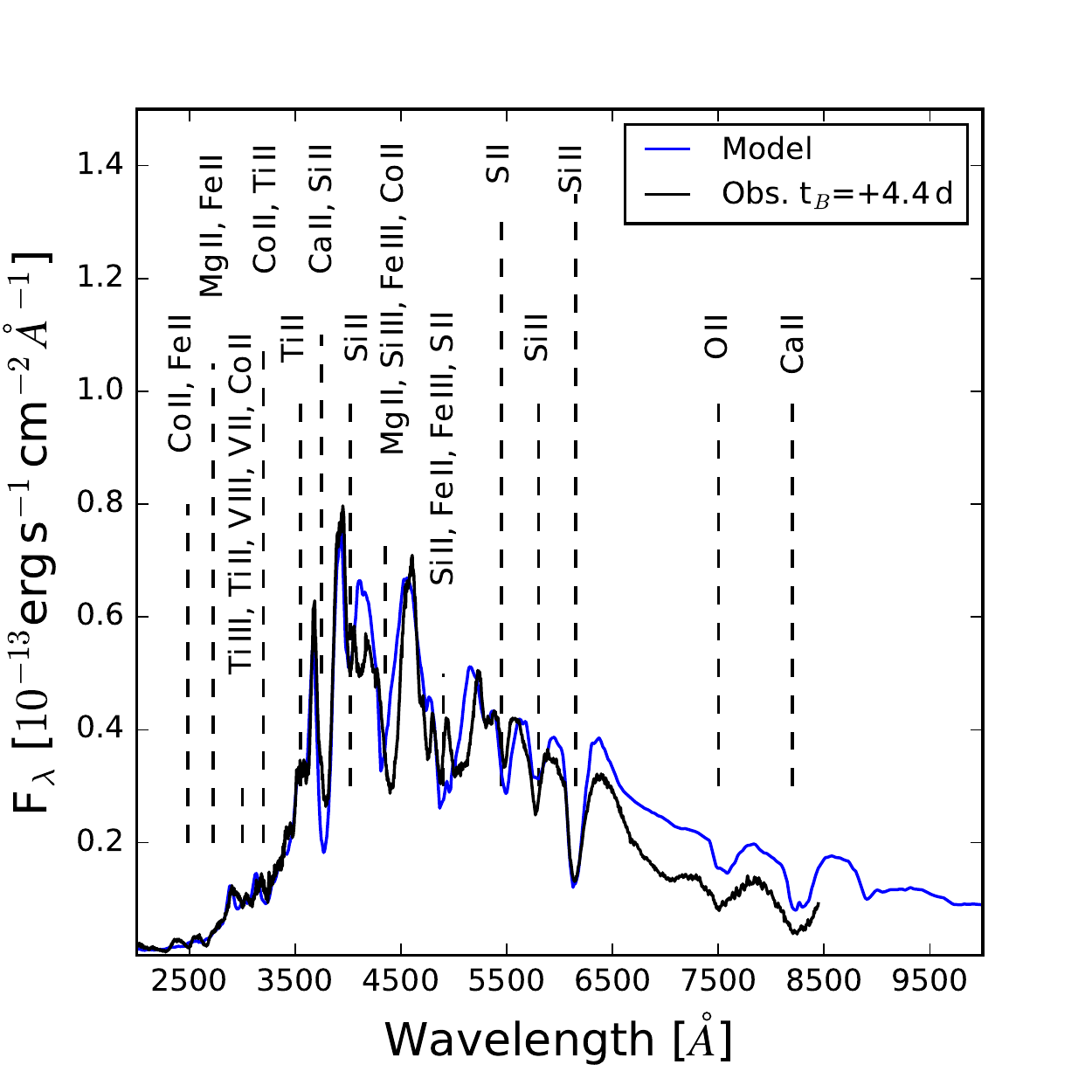} &\includegraphics[scale=0.7]{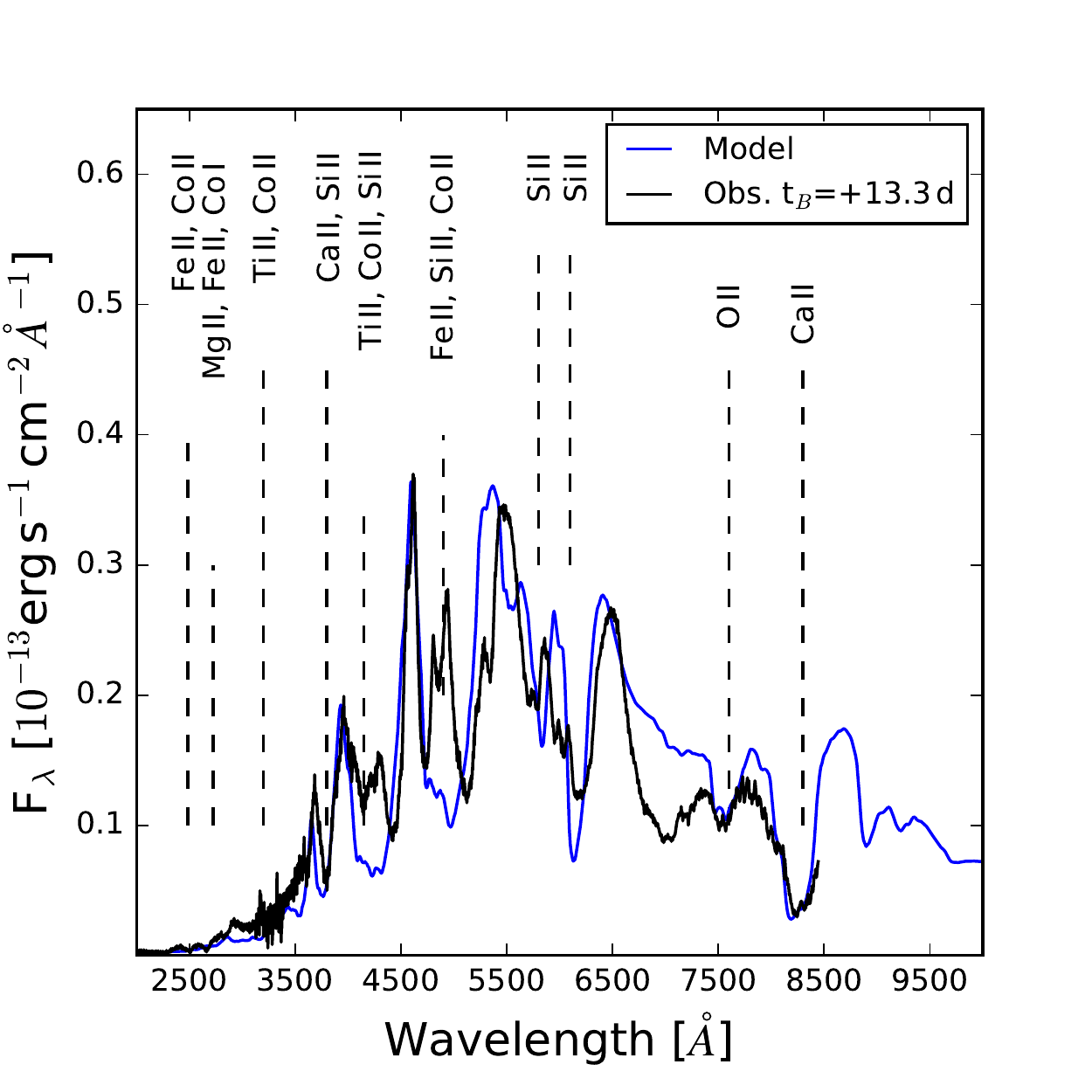} \\
 \end{array}$
 \caption{Spectrum (black) and model (blue) of SN\,2011iv at $-$5.5\,d (top left panel), 
 +0.4\,d (top right panel), +4.4\,d (bottom left panel), and +13.3\,d (bottom right panel) relative to $B$-band maximum. }
\label{fig:11ivdif}
\end{figure*}

The earliest spectrum, with UV data, of SN 2011iv was observed at $-$5.5\,d relative to $B$-band maximum.
The \vph\ at this epoch is 10200\,\kms. The main lines that contribute to each feature are identified in  the top left panel of Figure \ref{fig:11ivdif}.
In a SN,  photons emitted at the photosphere can only escape the ejecta when they reach an `available' frequency
at which they will not interact with any spectral lines. However, 
due to the relative expansion of the ejecta, and therefore Doppler overlapping of lines, the next available frequency for a photon 
to escape can be significantly redward relative to the frequency at which it was emitted from the photosphere. 
In the UV part of the spectrum this is a source of continuum opacity. This effect is known as line blanketing. 
Therefore,  setting the flux level in the UV is 
important for forming the optical spectra. 

The UV ($<3500\AA$) of the $-$5.5\,d spectrum consists of a blend of \CoII, \FeII, \MgII,   with some contributions from \NiII, \TiII, \TiIII\ and \VII.
The spectrum at wavelengths $<2500\AA$ is dominated by \CoII\ and \FeII\ lines, the strongest of these
 being \FeII\ $\lambda$2599 and  \CoII\ $\lambda$2580. In the wavelength region of 2500$-$3000\AA\ there
  is a blend of metal lines including \FeII\ $\lambda$2755, \CoII\ $\lambda$2663 and \NiII\  $\lambda$2510. The strongest lines in this region are the  \MgII\  resonance duplet $\lambda\lambda$2795,\,2802. Moving reward, the features are all dominated by the typical SNe~Ia 
 spectral lines \citep[see e.g ][]{Ashall14}.  The feature on the red side of the larger 4400\AA\ feature is caused by \SiIII\  \lam\lam4552,\,4567,\,4574 lines, and it should be noted that the 4800\AA\ feature is dominated by \SiII\ with only a small contribution from \FeIII.

\subsubsection{+0.4\,d}
The next spectrum with UV and optical data was the HST spectrum obtained close to $B$-band maximum.
The model at this epoch produces a good fit to the data, see the top right panel of Figure \ref{fig:11ivdif}.  
The \vph\ is 9500\,\kms, and the bolometric luminosity is log$_{10}$(L$_{bol}$)=42.95\,erg\,s$^{-1}$. 
Compared to the $-$5.5\,d model there has been some development of the spectrum,
most noticeably the shape of the flux in the UV. In this spectrum and model there
is a clear double absorption feature at 3100\AA,  and the shape of the 4400\AA\ and 4800\AA\
features have also changed. The change in the 4400\AA\ feature is due to a lack of \SiIII\ in the maximum light model and spectrum,
and the difference in the red part of the 4800\AA\ feature is due to an increase in the strength of \FeIII\ lines. 

In the UV the model fits the data exceptionally well. Once again the lines dominating this part of the spectrum and model
are blends of metals, as well as the \MgII\ resonance line duplet.  The double absorption feature at 3100\AA\ is dominated by weak \VII\ lines
\lam\lam 3093,3102, \TiII\ \lam 3349 and \TiIII\ \lam2984. The reduced flux between 3300-3800\AA\ compared to previous epochs is caused by a blend of \CoII\ lines.

\subsubsection{$+$4.4\,d}
The $+$4.4\,d HST spectrum and model are plotted in the bottom left panel of Figure \ref{fig:11ivdif}.
This model was 22.8 days after explosion. It is apparent at this epoch 
that the SN is much cooler. 
The model  produces a satisfactory fit, and the quality of the model in the UV is very good. 
The excess flux in the red is due to the model using the Schuster-Schwarzchild approximation, but since most of the strong lines are 
at wavelengths less than 6500\AA, this does not affect the results concerning line identification or abundances. 
In this model the 4800\AA\ feature consists of \SiII\, \FeII\ and \FeIII. 

\subsubsection{+13.3\,d}
The photospheric velocity of the +13.3\,d model is 6000\,\kms. This is  below $\sim$65\% of the 1.38\,$\Msun$\ ejecta. The fits to the data are 
worse at this epoch, as this is where there is a significant fraction of \Nifs, and therefore energy deposition, above the photosphere. However, 
the main absorption features are still seen in the models, and the flux in the UV is approximately correct, see the bottom right panel of Figure \ref{fig:11ivdif}. 

\subsubsection{Metalicity and metal content of SN~2011iv}
SNe Ia  with different progenitor metallicities will
produce varying amounts of \Nifs\ and Fe-group elements. These elements
 can be distributed differently throughout the ejecta. 
The UV region of a SN Ia spectrum is where the abundance of \Nifs\ and Fe-group elements can be best determined 
\citep[e.g][]{Lentz00,Walker12}. It has been theorised that an increase in progenitor metallically may affect the amount of \Nifs\
synthesised in SNe Ia, as a larger abundance of neutron-rich isotopes, such as $^{54}$Fe and $^{58}$Ni will be produced 
\citep{Iwamoto99,Timmes03,Bravo10}.
This will affect the shape of a SN Ia light curve \citep{Mazzali06}.  
The UV probes the area where metallically effects are also important, as a large number of metal line transitions are in this region.  
Photon packets that escape the SN ejecta in the UV must first be shifted to redder wavelengths furtherout in the ejecta before they can make a transition into the blue and escape the outer shell of the ejecta \citep{mazzali2000}.

\begin{figure}
\centering
\includegraphics[scale=0.7]{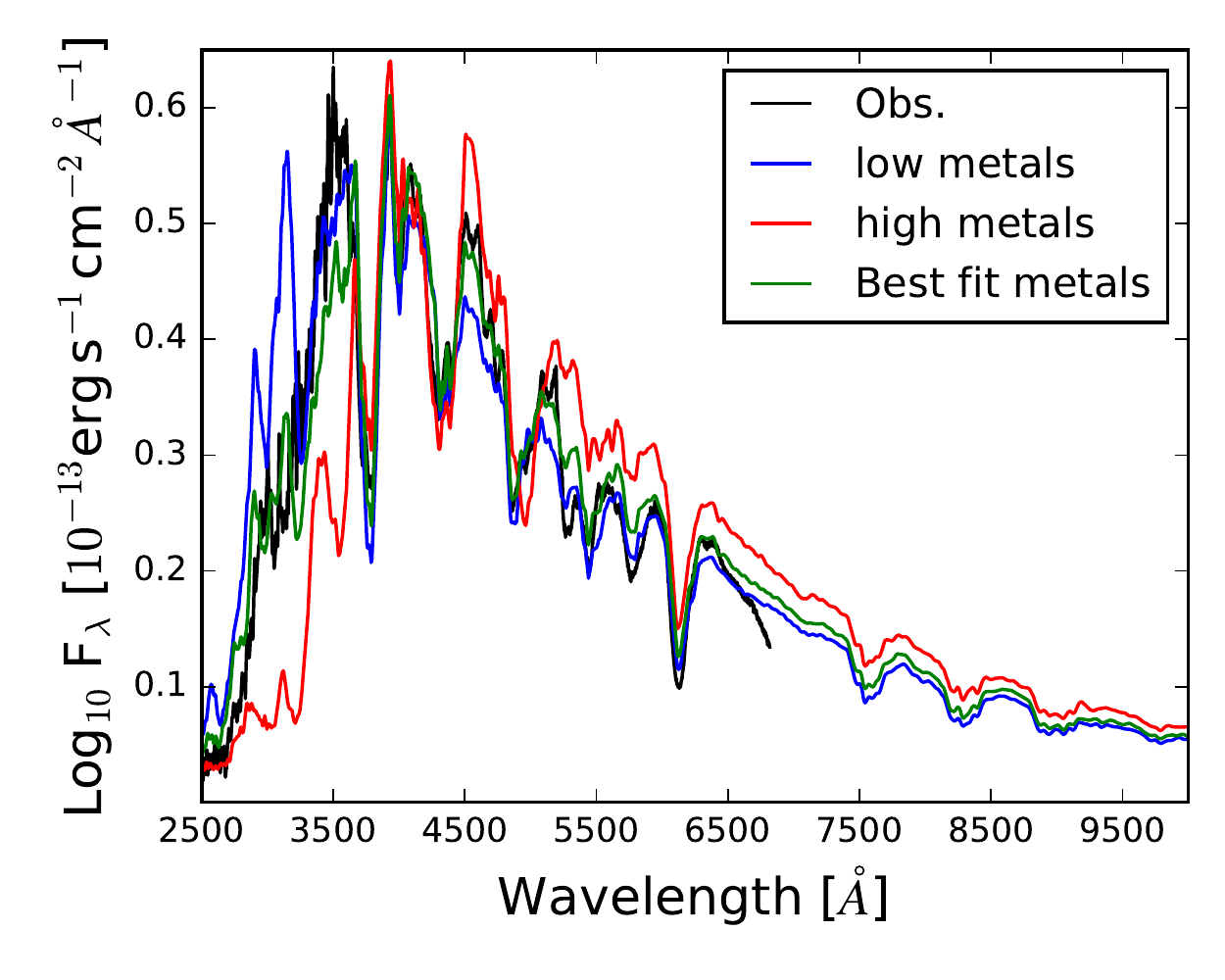}
\caption{The observation (black)   and three models with varying metal content of SN\,2011iv at  -5.5\,d. 
 The red model has a Fe abundance of 1.3\% above 15000\,\kms, the blue model has a Fe abundance of 0.03\% (i.e. a mass fraction 3$\times10^{-4}$) above 15000\,\kms,
and the green model is the best fit with an Fe abundance of 0.3\% above 15000\,\kms.}
\label{fig:metal}
\end{figure}

The early-time UV flux of a SN Ia ejecta is very dependent on the metal content. The UV flux level 
plays a major part in forming the optical part of a SN spectrum. For example, a higher metal content 
increases the amount of absorption in the UV, and the amount of line blanketing. This increases the back scattering 
rate of photon packets into the photosphere, which increases the photospheric temperature. 
In the outermost layers, the flux in the UV is directly linked to the metalicity of the progenitor system, as nucleosythesis is
expected to play a minor role. The spectrum obtained at $-$5.5\,d
relative to maximum is the earliest UV observation of SN 2011iv, see Figure \ref{fig:metal}. The photospheric velocity at this 
 shell is 10200\,\kms, and there is $\sim$0.25\,$\Msun$ of ejecta above this photosphere.  
 However, the amount of ejecta mass probed is a function of wavelength, and before maximum light 
 the flux in the $u$-band probes the outermost 0.06\,$\Msun$ of ejecta, with the outermost  $\sim$0.02\,$\Msun$
 of ejecta being probed at $-$5.5\,d\footnote{We note that these values will be dependent on the explosion model used.}\citep{Gall17}.
 Therefore our models are probing the outermost layers of the SN explosion where 
the composition and metallically of the progenitor system can affect the observed spectra. 
 
We have produced three one-zone models with varying metal abundances at $-$5.5\,d (see Figure \ref{fig:metal}). The best fit model
has an Fe abundance of 0.3\% above 15000\,\kms, and a \Nifs\ abundance of $\sim$1.5\%.  
The red model in Figure \ref{fig:metal} has a Fe abundance of 1.3\% above 15000\,\kms\ and the  blue model has a Fe abundance of 0.03\%.
The outer layers have metalicity larger than solar \footnote{These results could change depending on the underlying density profile used, and we note our results 
will be less sensitive at velocities over $\sim19000$\,\kms.} \citep{Asplund09}. 
If this high velocity, large metal abundance is not primordial, it is still likely to be linked to the 
metalicity of the progenitor WD. It has been demonstrated by many groups (see, e.g., \citet{Hoflich98,Iwamoto99}) that 
a small increase in progenitor metalicity can dramatically increase the production of $^{54}$Fe in the outer layers.
This is because  metalicity mainly affects the initial CNO abundances of a star, and therefore
the proton-to-nucleon ratio ($Y_{e}$) decreases with increasing metalicity \citep{Hoflich98}. 
Conversely, in the low-velocity central regions of a SN Ia the temperatures are high enough that $Y_{e}$ is determined by electron capture. Therefore, the large Fe abundance in the outer layers of SN\,2011iv  implies that the metalicity of the progenitor WD was super-solar. This is in agreement with what could be expected from a SN in a giant elliptical galaxy, such as NGC\,1404.

\subsection{Comparison between the photospheric results of SN~2007on and SN~2011iv}
The early time models of SN~2007on and SN~2011iv highlight the fact that around maximum they are similar objects, which
can be identified by similar spectral lines. 
However they do show some differences. For instance the rise time and bolometric luminosity of SN\,2011iv 
($t_{rise}$=18.4\,d, log$_{10}$(L$_{max}$)=42.95\,erg\,s$^{-1}$ ) is larger than the equivalent values for SN~2007on
($t_{rise}$=16.8\,d, log$_{10}$(L$_{max}$)=42.60\,erg\,s$^{-1}$). Furthermore, the spectra of SN\,2011iv tend to contain more
doubly ionised species (\eg \FeIII) with respect to SN~2007on, although both have significantly less  than a `normal' SNe Ia.
To first order the difference in rise time and light curve shape of the objects is due to an opacity effect (as opacity is the 
main driver of light curve shape) where the more luminous SN~2011iv produces more Fe through \Nifs\ decay and 
therefore has a broader light curve (as line opacity dominates in SNe Ia, and Fe-group line contribute to the opacity by an order
of magnitude more than those from intermediate mass elements (IME) \citep{Mazzali07}). In section 5 we examine the abundances obtained from the
spectral modelling and  further quantify the main reason for the difference between SN~2007on and SN~2011iv.
 
The other notable difference between the SNe is that the best models are produced with different density profiles. 
The preferred density for SN~2007on (W7) has more mass at higher velocities than the best density for SN~2011iv (W7e0.7).
 This implies that if SN\,2007on is a Ch-mass explosion, it will have less mass in the inner layers relative to SN\,2011iv,
which agrees with the conclusions of \citet{Gall17} who determine that SN\,2007on had a lower $\rho_{c}$ than SN\,2011iv.
 Furthermore, \citet{Gall17} found that around maximum light the Doppler velocity of 
 the  \SiII\ $\lambda$6355 feature is larger in SN\,2007on than SN\,2011iv; this is in 
 agreement with the results here which show that SN\,2007on has a higher \KE\ than SN\,2011iv. 
Interestingly the least luminous of the two SNe was
produced from the more energetic explosion.   
This highlights the peculiarities of SN~2007on and hints that these objects may come from different progenitor scenarios or explosion mechanisms. 

\section{Nebular phase}
The inner layers of SN\,2007on maybe formed of two separate components. SN\,2007on is therefore 
the subject of a companion paper \citet{Mazzali17}. Hence we do not discuss the nebular spectra of SN\,2007on here.

\subsection{SN 2011iv}
The nebular spectrum of SN\,2011iv was modelled using a non-local thermodynamic
equilibrium code that was described in a series of papers
\citep[\eg][]{Mazzali01}. In order to test the general properties of the SN we
started with a one-zone model where the abundances are kept constant inside a
sphere with limiting radius determined by the velocity of the emission lines.
The epoch of the spectrum was 276 rest-frame days after explosion, or \ab260\,d
after $B$-band maximum (observed). The boundary velocity that led to best fits
was selected as 9000\,\kms. We built up the model trying to match the strongest
features caused by the [\FeII] and [\FeIII] emission lines (see for
example \citet{Mazzali07}). The spectrum is powered mainly by the decay of
\Cofs, but the decay of \Nifs\ also contributes. We found that in order to match
the flux emitted in most of the strongest optical lines, a \Nifs\ mass of
0.41\,$\Msun$ is required. The [\FeIII]-dominated emission near 4700\AA\ is
suppressed relative to the [\FeII]-dominated emission near 4300\AA. To produce
the rather low observed ratio, \ab1.5, of the two strongest Fe lines, a
significant amount of stable iron group elements (0.25\,$\Msun$ of stable iron, and 0.09\,
$\Msun$ of stable nickel) is necessary. These stable iron group elements
elements contribute to cooling but not to heating, thereby reducing the
ionization state. The stable nickel gives rise to an emission near 7300\AA,
which matches the observed feature in flux but not in wavelength as the emission
line is affected by significant blue-shift.  The model has a total mass
\ab0.8$\,\Msun$ inside the boundary velocity 
(The nebular phase 1-zone model contains 0.05\,$\Msun$ of IMEs). 
This is consistent with a Ch-mass explosion of a C-O WD.

In order to improve on these results, and to verify the consistency between the
nebular modelling and the early time models, we also modelled the nebular
spectrum using the abundance tomography approach. We extended the density and
compositions used in section 3.2 to verify whether they are compatible and
viable at lower velocities. In practice, we modified the abundances at
velocities below 6000\,\kms, preserving the density distribution derived through
early-time modelling. We find that our stratified model is consistent with the
observations, and indeed yields better results than the one-zone model, as 
expected.  We also find, in agreement with the one-zone model, that the
stable iron group species are highly abundant in the innermost layers.  In fact,
they dominate the composition below 5000\,\kms. In particular, stable iron
dominates between 3000\,\kms and 4500\,\kms, and stable nickel dominates below
2000\,\kms.  The total masses we derive were, 0.31\,$\Msun$ of \Nifs, 0.27\,$\Msun$
of stable Fe, and 0.08\,$\Msun$ of stable Ni. The total mass of iron group
elements is 0.66\,$\Msun$, which is consistent with a moderately luminous SN Ia 
\citep{Mazzali08}. Other emission lines  produced by the model
are the \NaI-D feature near 5900\AA\, \CaII] emission near 7200\AA, as
well as [\FeII] near 8500\AA, and [\NiII] near 7400\AA. The observed [\NiII]
line is somewhat blue-shifted relative to its rest wavelength, as noticed in
\citet{Mazzali17}. A weak [\SII] line is also present near 4000\AA.

\begin{figure}
\centering
\includegraphics[scale=0.7]{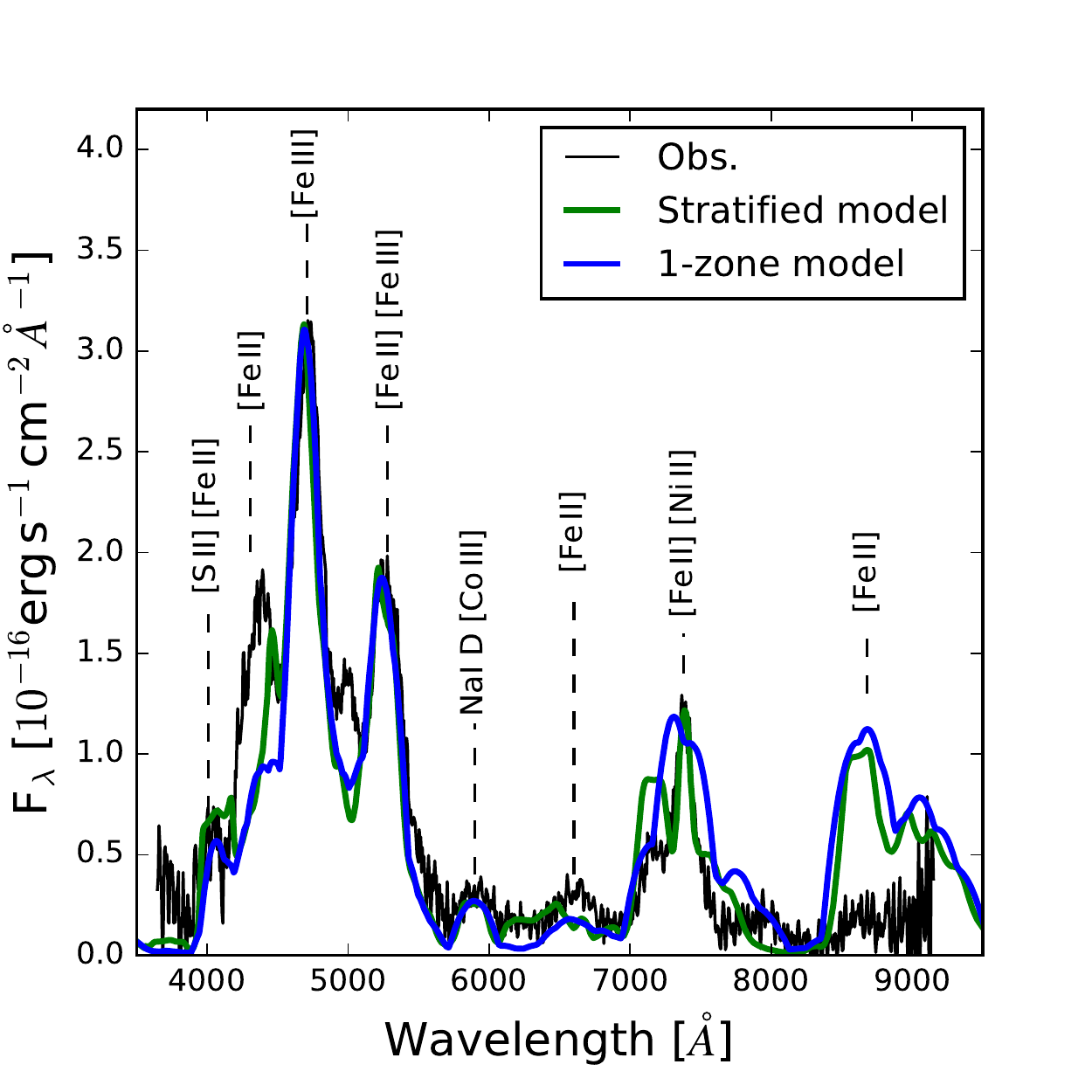}
\caption{Nebular phase spectra (black), the one-zone model (blue) and the stratified model (green) of SN\,2011iv at $+$276\,d after explosion.}
\label{fig:neb1zone}
\end{figure}

\section{Abundance Stratification}
Having produced spectral models of the SNe we now examine their abundances in mass and velocity space.  

\subsection{SN~2007on}
\begin{figure*}
\centering
\includegraphics[scale=0.75]{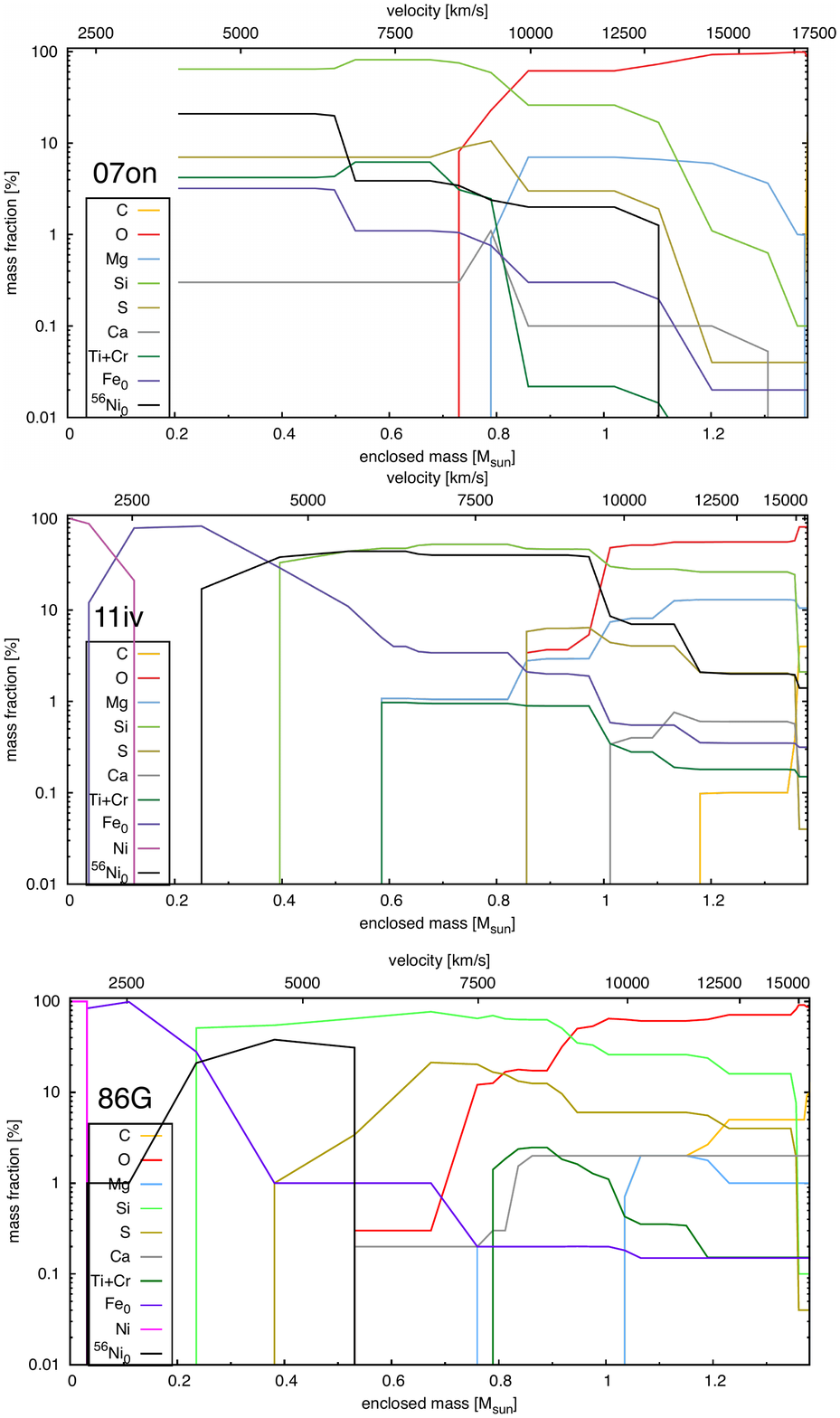}
\caption{{\it Top:} The final abundances of the main elements of SN\,2007on  derived from spectral modelling, using
 the W7 density profile. 
{\it Middle:} The final abundances of the main elements  of SN\,2011iv derived from spectral modelling, using the W7e0.7.
{\it Bottom:} The abundances of SN~1986G calculated in \citet{Ashall16b}. These values were derived using the W7e0.7 density profile. The abundances are given at time t=0s, $^{56}Ni_{0}$ denotes the abundance 
of \Nifs\ at time t=0s relative to explosion, and Fe$_{0}$ is the stable Fe abundance at t=0s relative to explosion.}
\label{fig:tomo}
\end{figure*}

The abundance distribution for the photospheric models of SN\,2007on is presented in the top panel of
 Figure \ref{fig:tomo}.
The oxygen zone dominates down to \ab10000\,\kms, and the IMEs zone dominates the ejecta
down to the lowest photospheric velocity of 3900\,\kms. There is \Nifs\ located up to \ab12500\,\kms, but in a
 much smaller abundance than SN\,2011iv. The abundance distribution is what would be expected of 
 a thermonuclear explosion of a WD. It has a chemically layer structure which is consistent with a moderately sub-luminous delayed detonation models of Ch-mass WDs, for example see \citet{Hoflich17}. 
 
It is not clear how the  one component model of the outer layers is connected to the inner two component model. 
The problem deviates from 1D, and it would not make sense to reconstruct the abundance distribution in the innermost layers using the standard abundance tomography approach. 
However, at early times these inner layers are below the optically thick photosphere, hence the number of components in the low velocity layers is not important. 

 Although the whole abundance distribution of SN~2007on cannot be computed,
 we can examine the
 integrated abundance from the early time models in this work and the nebular phase models 
 from \citet{Mazzali17}. 
 The nebular phase models are sensitive out to 4000\,\kms,
 therefore the integrated abundances from this work are used above 4000\,\kms, and those from  
 \citet{Mazzali17} below this. The abundances are given in Table \ref{table:07onabund}.
 The total ejecta mass is 1.42\,$\Msun$, with 0.5\,$\Msun$ of unburnt material, 0.61\,$\Msun$ of IMEs,
 0.25\,$\Msun$ of \Nifs\ and 0.06$\Msun$ stable nuclear statistical equilibrium (NSE) elements.
 The \Nifs\ mass is similar to that derived from the 
 peak of the light curve \citep{Gall17}. 

With the integrated masses obtained from the modelling, the
\KE\ of the explosion can be derived, using the formula  
\begin{equation}
E_{k}=[1.56M(^{56}Ni)+1.74M(sNSE)+1.24M(IME)-E_{BE}]10^{51}\,erg
\label{eq:KE}
\end{equation}
\citep{Woosley07}, where $E_{BE}$ is the binding energy of the progenitor 
WD, M(\Nifs) is the \Nifs\ mass, M(sNSE) is the mass of stable NSE elements,
and M(IME) is the total IME mass in the ejecta. 
For a standard energy explosion of Ch-mass WD, $E_{BE}=0.46 \times 10 ^{51}$\,erg.

Using equation (1) a \KE\  of  (1.24-$E_{BE})\times10^{51}$\,erg is obtained,
 where $E_{BE}$ varies depending on the explosion scenario. 
 If the normal $E_{BE}$=0.46$\times10^{51}$\,erg is used, a total \KE\ =0.78$\times10^{51}$\,erg is obtained,
 which is inconsistent with the input model with 1.3$\times10^{51}$\,erg. 
 To achieve a much lower $E_{BE}$  requires a low $\rho_{c}$. 
 A low $\rho_{c}$ is in line with the nebular spectral models 
 (see \citet{Mazzali17}), which show that SN~2007on had a low mass of stable NSE elements (0.06\,$\Msun$)\footnote{A low $\rho_{c}$ implies a low electron capture rate and pushes
  burning away from stable NSE elements.}. 
 This is much lower than other SNe Ia (see \citet{Stehle05,Ashall16b}).
It is not apparent whether a Ch-mass WD can produce such a low mass of stable NSE elements. 
In the  collision scenario of two 0.7\,$\Msun$ WDs the $\rho_{c}$ and $E_{BE}$ will be lower. 
 $E_{BE}$ for a non-rotating WD with mass 0.7\,$\Msun$ is 0.02$\times10^{51}$\,erg \citep{Yoon05}. 
Therefore, the  collision scenario gives consistency between the  \KE\ (1.2$\times10^{51}$\,erg) 
 calculated from the nucleosynthesis yield,  and the \KE\ calculated with the
 input density profile (1.3$\times10^{51}$\,erg). 
 
As noted previously, SN\,2007on shows two late-time nebular components, both of these components have a similar ejecta mass 
$\sim0.1\,\Msun$ and a smiliar \Nifs\ mass of \ab0.085\,$\Msun$, but a different ionisation state caused by different amounts of 
stable NSE elements, which act as coolants. Two of the scenarios which could produce two late-time components. 
These are the  collision of two WDs or an off-centre ignition of a WD in the delayed denotation scenario.

In an off-centre ignition scenario the detonation phase starts away from the centre of the WD, which causes 
a `ring'-like structure of 
\Nifs\ around the central core, with a `blob' of \Nifs\ where detonation started \citep[see \eg][]{Fesen07}.  
Within the delayed detonation scenario, \Nifs\  is produced during 
both the deflagration and detonation burning phases.  As we will discuss
in Section 7, the production is dominated by the  detonation
phase in normal-bright SNe Ia, and by the deflagration phase in  sub-luminous objects. 
For the transitional
SNe Ia such as SN\,2007on and SN\,2011iv, the best fitting delayed detonation models produce
about equal amounts of \Nifs\ in each phase, and the point of detonation occurs at \ab7000\,\kms \citep[see ][]{Gall17}.  
 The corresponding off-center
2D simulation has been applied to S-Andromeda and it was found that a similar amount of \Nifs\ was produced
in each phase, and
off-set with an average velocity of 6000\,\kms\ \citep{Fesen07}. Detailed tuning of 
parameters is beyond the scope of this study, but results in the literature make off-center delayed detonations a 
viable option because components with similar amount of \Nifs\  and a slightly larger separation have been seen.

The abundance distribution of the outer layers of SN\,2007on 
is not in disagreement with an off-centre explosion, however most signatures of an off-centre
explosion can be seen in the nebular 
phase and we cannot draw any strong conclusions on this with just the early time data. 
For a more detailed discussion see \citet{Mazzali17}. 
Other aspects, such as the colour curve and the light curve, of SN\,2007on
also match the delayed denotation scenario  \citep{Gall17,Hoflich17}.

 However, the two things that it seems cannot be explained by the delayed denotation
  scenario are the lack of stable NSE elements produced in the ejecta (i.e. can a single Ch-mass WD have a low enough $\rho_{c}$ to
  produce the very small abundance of stable NSE elements seen in SN\,2007on?), 
 and  the similar ejecta masses of each nebulae component.
The  collision of two \ab0.7\,$\Msun$ WDs can explain the lack of stable NSE elements, due to the lower $\rho_{c}$. 
It also naturally explains the equal mass of each nebula. Additionally, \ab0.7$\Msun$ is the 
typical mass of WD in the Universe \citep{Kepler17}. 
The collisions of two WDs are likely to be rare, explaining 
why only a few SNe Ia have been claimed to have double peaked nebular lines \citep{Dong15}.

\begin{table}
 \centering
 \caption{ The integrated abundances of SN\,2007on. The abundances below 4000\,\kms\ are taken from \citet{Mazzali17}, and the abundances above 4000\,\kms\ are taken form this work.
 The errors on the masses are $\pm$25\%. }
 \begin{minipage}{250mm}
  \begin{tabular}{cccc}
  \hline
Material&v$<$4000\,\kms&v$>$4000\,\kms&total\\ 
&$\Msun$&$\Msun$&$\Msun$\\
  \hline
unburnt (C+O)\footnote{Some of this oxygen will have been produced through burning.}&0.00&0.50&0.50\\
IME&0.01&0.60&0.61\\
\Nifs&0.17&0.08&0.25\\
stable NSE&0.02&0.04&0.06\\
  \hline
Total&0.2&1.22&1.42\\
  \hline
\end{tabular}
 \end{minipage}
\label{table:07onabund}
\end{table}

\subsection{SN~2011iv}
The abundance distribution of SN\,2011iv as a function of mass and velocity has been reconstructed (see the middle panel in 
Figure \ref{fig:tomo}).
The nucleosynthesis follows what would be expected of a thermonuclear runaway reaction. However, the unburnt and IME zones are much 
larger in SN\,2011iv than in a more luminous SN such as SN\,2011fe. This indicates that the burning was less efficient.
The oxygen zone dominates down to $\sim$9500\,\kms, and there are signs of a small amount ($<1\%$) of carbon in the outermost layers.
The abundance distribution indicates that there is  mixing in the ejecta, with \Nifs\ and Ti+Cr located in the outermost velocities ( >15000\,\kms).
The IME zone dominates from 9500\,\kms\ down to 6000\,\kms, and the inner-most layers, below 6000\,\kms,  mainly consist of NSE elements. 

As discussed in Section 3.2.5, a SN Ia with a large progenitor metalicity will have a lower $Y_{e}$ in the outer layers. This low $Y_{e}$ provides a large $^{54}$Fe production in these high-velocity layers. 
Usually, the Fe production also decreases in intermediate velocity layers, and peaks in abundance
 in the central densest regions. In SN\,2011iv there is a larger than average Fe abundance in the outer layers, which has been linked back to the progenitor metalicity (see Section 3.2.5). However, there is not a decreasing Fe abundance in the intermediate velocity regions, see Figure \ref{fig:tomo}. In the companion paper \citep{Mazzali17} it was found that SN\,2011iv could have come from an off-centre ignition of a WD. Therefore, we speculate that the increase in Fe abundance in the intermediate velocity regions is caused by the off-centre ignition of the WD.

The final integrated abundances are presented in Table \ref{table:intabun}. SN\,2011iv produced 
0.24\,$\Msun$ of unburnt (C+O) material, 0.48\,$\Msun$ of IMEs, 0.35\,$\Msun$ of stable NSE 
and 0.31\,$\Msun$ of \Nifs. 
This is in line with the two component models of \citet{Mazzali17} which find that SN\,2011iv has 0.32\,$\Msun$ of \Nifs. 
Putting the integrated masses obtained from the modelling into equation \ref{eq:KE}, the
kinetic energy of  SN\,2011iv calculated from the abundances ($1.2\pm0.3\times10^{51}$\,erg)
\footnote{For a low-energy explosion of Chandrasekhar mass WD $E_{BE}=0.49 \times 10 ^{51}$\,erg.}
is similar to the \KE\  of the input density profile, W7e0.7 ($0.9\pm0.2\times10^{51}$\,erg). 
Given the error on the abundances and \KE\ of the density profile, there 
is good consistency between the \KE\ derived through abundances and the \KE\ obtained from the input density profile. 
We note that the \Nifs\ mass ratio (0.81) of  SN\,2007on to SN\,2011iv (0.25/0.31\,$\Msun$) obtained in this work
  is 
remarkably similar to that derived from the delayed detonation models used in  \citet{Hoflich17} and in appendix D
 of \citet{Gall17} who obtain a ratio of 0.86 (0.32\,$\Msun$/0.37\,$\Msun$). The absolute difference in \Nifs\ masses is due to the slightly larger distance modulus used in \citet{Hoflich17} (see Section 2).

The abundance distribution indicates that SN\,2011iv was consistent with a  Ch-mass delayed detonation explosion of a C-O WD. In this scenario, 
the explosion  had a low deflagration to detonation $\rho_{tr}$, which caused the WD to partially unbind and 
the burning in the detonation phase to be less efficient, producing more IMEs.   The fact that there is \Nifs\ located further out in the ejecta
 indicates that there was some mixing, possibly during the deflagration phase. 
The lack of \Nifs\ below 2500\,\kms\ is in line with explosion models. 
A higher $\rho_{c}$ will increase the electron capture rate and produce stable NSE elements in the innermost layers \citep{Hoflich17}. 
As the \Nifs\ is located further out in the ejecta the mean free path of the photons will be longer, 
causing the rise of the SN light curve to be faster, which matches the observations. 
These results add to the growing evidence that the $\rho_{c}$ of SN\,2011iv was higher than standard SNe Ia.
The results of this abundance distribution plot also complement the results of \citep{Mazzali17} who determine that SN\,2011iv was an off-centre explosion.

\begin{table}
 \centering
 \caption{Integrated abundances from the full abundance tomography modelling of SN 2011iv.
  The errors on the masses are $\pm$25\%.}
  \begin{tabular}{cc}
  \hline
   Element&W7e0.7\\
   &$M_{\sun}$\\
  \hline
C&$<$0.01\\
O&0.24\\
Mg&0.05\\
Si&0.39\\
S&0.03\\
Ca&0.01\\
Ti+Cr&$<$0.01\\
Fe&0.27\\
$^{56}Ni$&0.31\\
Ni&0.08\\
\hline
$M_{tot}$&1.38\\
\hline
\end{tabular}
\label{table:intabun}
\end{table}

\subsection{Comparison of abundance stratification}
In delayed detonation explosion models  fainter SNe Ia have a smaller $\rho_{tr}$ and a longer 
deflagration phase, which results in greater pre-expansion during deflagration and less effective burning during detonation, thus producing less \Nifs.
 The abundance stratification of three transitional SNe Ia (SNe~2007on, 2011iv, and 1986G) is presented in Figure \ref{fig:tomo}.
If the results of these SNe~Ia are interpreted in terms of delayed denotation explosions, 
SN~2011iv synthesised the most \Nifs\ (0.31\,$\Msun$) and
therefore had the largest $\rho_{tr}$. The pre-expansion during the deflagration phase
 partially unbound the WD, therefore the burning during the detonation phase  was less effective than in a normal SN Ia. 
  SN~2007on had a $\rho_{tr}$ between that of SN~2011iv and 
 SN~1986G. SN~2007on had a longer deflagration phase than SN~2011iv, which further unbound the WD
  producing less effective burning during the detonation. 
Hence, SN~2007on
synthesised an intermediate amount of \Nifs. 
SN~1986G had the lowest $\rho_{tr}$ and produced the least amount of \Nifs\ with virtually none being located at high velocity.  However, this picture does not take second order parameters such as 
$\rho_{c}$ of the WD into account. 
 
 In fact in the delayed detonation scenario, SNe~1986G and 2011iv can be thought of as very similar explosions, with different $\rho_{tr}$. 
SN~1986G had lower photospheric velocities,  a larger abundance of unburnt material (0.34\,$\Msun$), and a \KE\ of 1.0$\pm$0.2$\times10^{51}$erg\footnote{\KE\  calculated from integrated abundances.}.
On the other hand, SN~2011iv had larger photospheric velocities, 
 a smaller abundance of unburnt material (0.24\,$\Msun$) and a \KE\ of 1.2$\pm$0.3$\times10^{51}$erg$^{11}$. 
In the less luminous SN~1986G more oxygen and carbon remained unburnt
 and did not contribute to the energy production, hence the \vph\ and \KE\ were lower. 
 More similarities between both SNe are the facts that their models favoured the same density profile, 
 they both  had a high $\rho_{c}$, as well as an  off-centre \Nifs\ distribution. 
 This is evidence that SNe~2011iv and 1986G came from very similar progenitor systems. 

It may actually be the case that SNe~2007on and 1986G were also similar explosions.
 The total mass of Fe-group elements for SN~2007on was 0.31\,$\Msun$, and for SN~1986G was 0.34\,$\Msun$.  In the delayed detonation scenario, this indicates
that their $\rho_{tr}$ were similar, but that $\rho_{c}$ is the main difference between these objects.
The high $\rho_{c}$  in SN\,1986G increases the amount of stable NSE elements produced at the expense of \Nifs. 
Therefore, we have a situation where SN\,2011iv and SN\,1986G have similar $\rho_{c}$ and different $\rho_{tr}$, 
but SN\,2007on and SN\,1986G have different $\rho_{c}$ and similar $\rho_{tr}$. This suggets that they may all 
come from a similar progenitor scenario, and accounting for both parameters ($\rho_{c}$ and $\rho_{tr}$)
 is important if these transitional SNe Ia are to be used as distance indicators.
 
However, comparing these objects in terms of just $\rho_{tr}$ and $\rho_{c}$ is an over simplification. 
There could be other factors which differ between them, including the points of ignition. 
\citet{Mazzali17} demonstrated that SN~2011iv had an off-centre ignition, whereas \citet{Ashall16b} found no evidence for SN~1986G
igniting off-centre. It could also be the case that SN~2007on had an extremely off-centre ignition, and it should be noted that 
the outer layers of SN~2007on are consistent with a delayed detonation Ch-mass explosion. 
Although the collision of two WDs with a total ejecta mass near the Ch-mass is also a viable scenario \citep{Mazzali17}.

\section{Bolometric Light curves}
In the spirit of abundance tomography, we built a
bolometric light curve to test whether the model of SN~2011iv may be compatible with the
observations. 
The light curve was based on the density and abundance distributions derived above.
We used a well-tested Montecarlo code, which starts with the same description of
gamma-ray and positron emission/deposition as in the nebular phase code, and 
follows optical photons as they propagate through the SN ejecta using the grey
approximation. The code was first described by \citet{Cappellaro97}, then improved upon
by \citet{mazzali2000}, and it has been used several times in the study of both
SNe\,Ia \citep[\eg][]{Ashall16b} and SNe\,Ib/c \citep[e.g.,][]{Mazzali13}.

\begin{figure}
\centering
\includegraphics[scale=0.5]{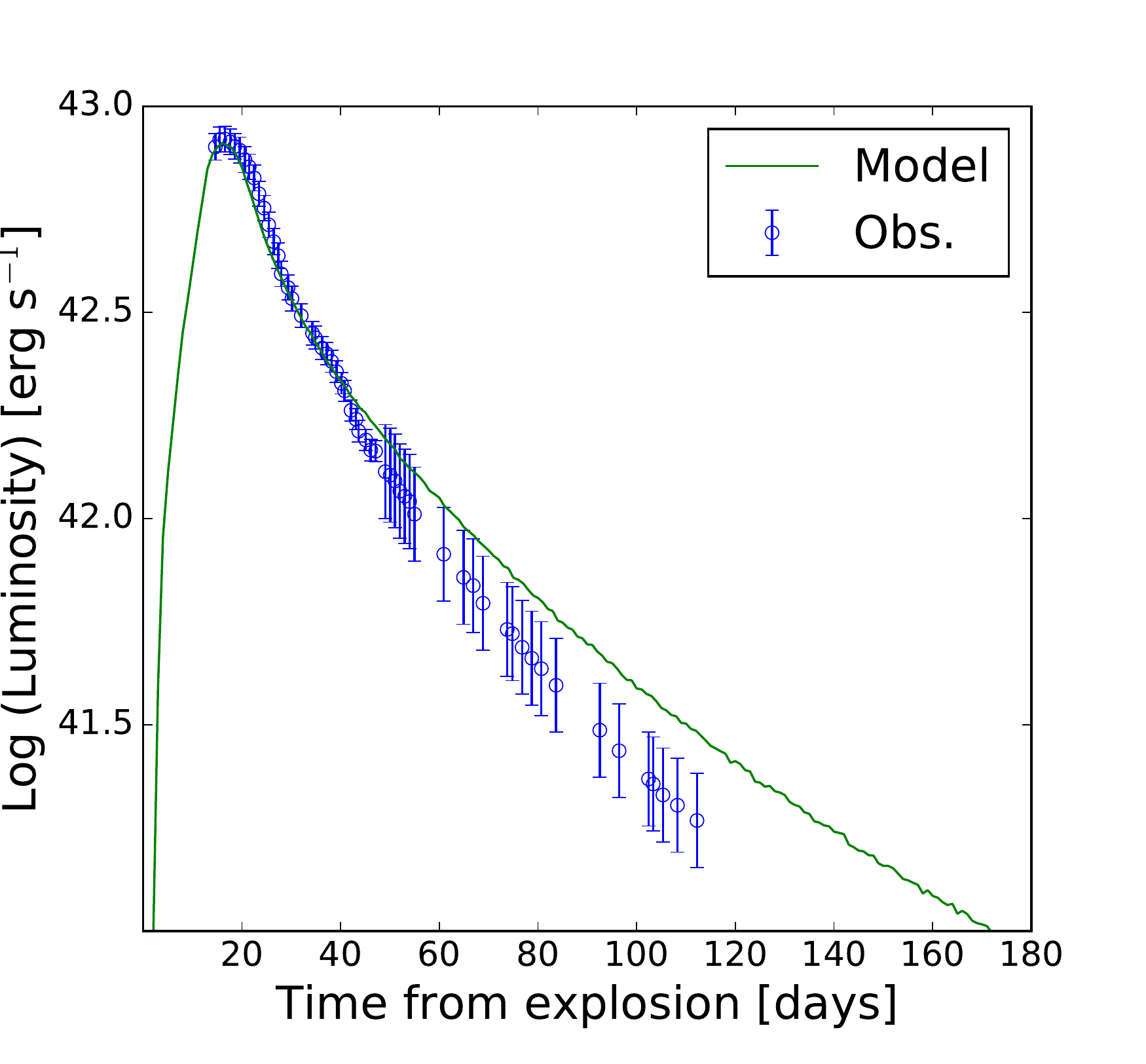}
\caption{ The observed (blue) and modelled (green) bolometric light curve of SN\,2011iv. The observed bolometric light curve of
SN~2011iv was taken from \citet{Gall17}, and constructed using \textsc{SNooPY} \citep{Burns11,Burns14}. The near IR photometric coverage of SN\,2011iv after 40 days past maximum light was poor, hence we adopted a standard flux error after this point of 30\%, which could be this missing contribution (see \citet{Gall17} Figure 11).  There could also be up to 10\% flux in the MIR at these epochs. }
\label{fig:BolLC}
\end{figure}

The synthetic light curve (see Figure \ref{fig:BolLC}) reaches maximum light $\sim$18\,d after explosion at
M(Bol)$_{Max}\sim-18.6$\,mag, or $\log(L) = $42.91\,erg cm$^{-2}$
s$^{-1}$. It matches the observed light curve very well both around peak and in
the  early decline phase, up to about day 40. Thereafter, the observed light
curve declines  more rapidly. As the light curve also reproduces the
observations at day 270 (this is by definition as the same code that is used to
compute the light curve is used to compute the deposition for the nebular
spectrum), due to a lack of data, it may be possible that the near-infrared and mid-infrared flux is not properly accounted
for in the construction of the  pseudo-bolometric light curve.

\section{Properties of transitional SNe Ia models}
One issue when comparing transitional and sub-luminous SNe Ia is the degeneracy 
in \Dm\ \citep{Phillips12,Burns14,Gall17}.
SNe Ia with \Dm$>1.7$\,mag  suffer from the fact that two SNe Ia with different light curve shapes can have 
the same value of \Dm.    
As opacity is the main driver of SNe Ia early time light curve shape \citep{Mazzali07},
fainter SNe Ia have faster light curves than standard SNe Ia.
This is because line opacity dominates in SNe Ia \citep{Pauldrach96},
and Fe-group elements contribute to the opacity by a factor of 10 more than intermediate mass elements.
Therefore a SN Ia 
with a smaller \Nifs\ mass will have a low opacity and a fast evolving light curve 
(as \Nifs\ decays to $^{56}$Co and $^{56}$Fe).
In the most extreme cases the least luminous SNe Ia, 
with very little \Nifs, reach the \Cofs\ decay tail in the light curve 
before 15 days past $B$-band maximum.
 Therefore objects such as SN~2005ke and SN~2006mr would be directly comparable
if just \Dm\ was used.  \citet{Burns14} proposed a new method of characterising light curve shapes.
 This is the colour stretch parameter, $s_{BV}$. It is a dimensionless parameter 
defined to be the time difference between $B$-band maximum
and the epoch the $B-V$ colour reaches its maximum value, divided by 30 days. Ordering 
fast declining SNe~Ia by $s_{BV}$ breaks this degeneracy and places the SNe in order of peak 
luminosity, but it does not alter the fact  there
is still some variation between fast-declining SNe Ia. 
For example, SN 2011iv sits above the $s_{BV}$ vs. $M_{B}$ relation, 
and it does not answer the question why does SN\,2007on show double peaked emission in the nebular phase. 
SNe~2007on and 2011iv have  $s_{BV}$ values of  0.57$\pm$0.01 and 0.64$\pm$0.01, respectively \citep{Gall17}. 

\begin{figure*}
\centering
\includegraphics[scale=0.7]{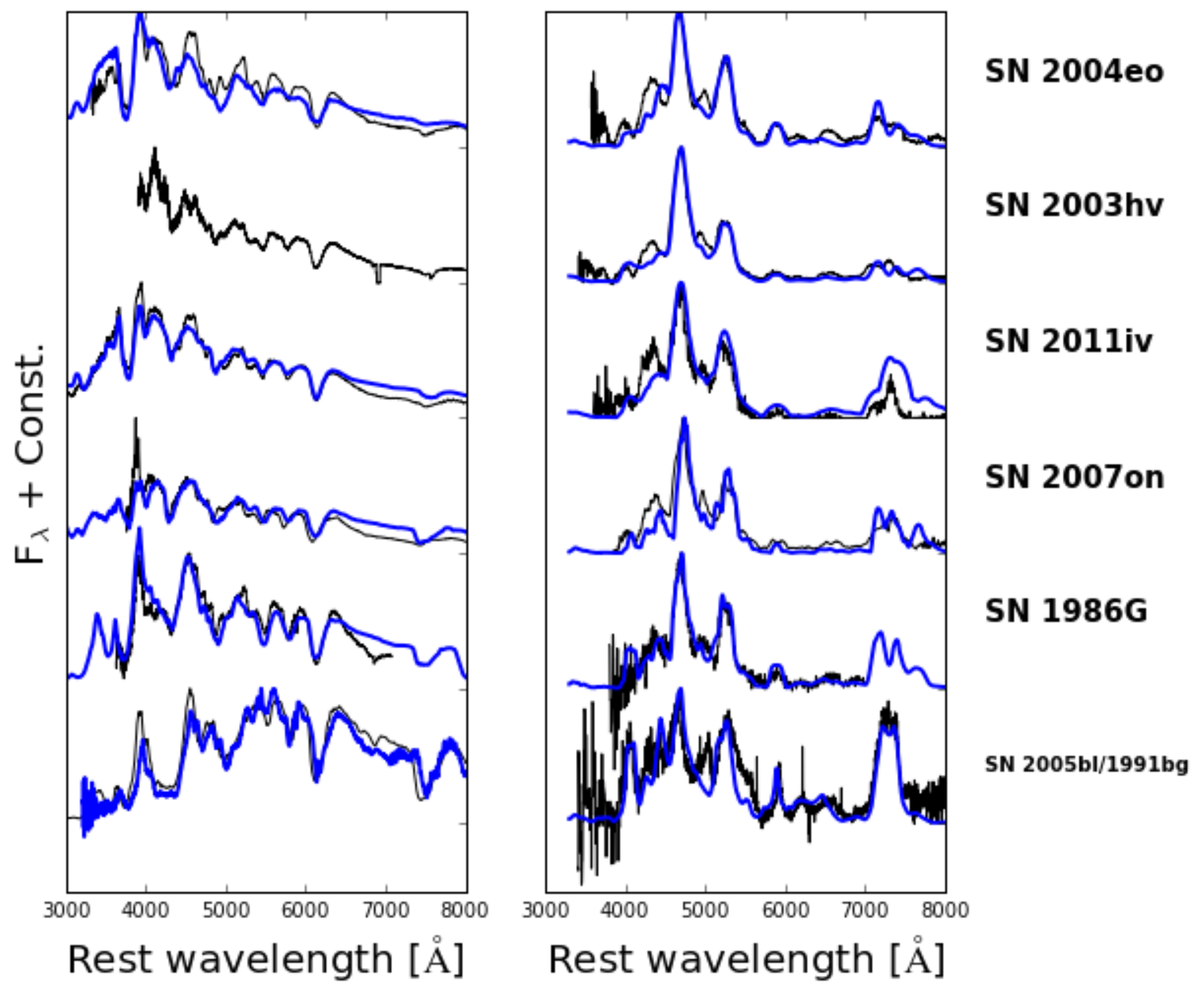}
\caption{Left: Maximum light spectra (black) and models (blue) of the SNe in Table \ref{table:transi}. No models have been made for the early time of SN\,2003hv or SN\,1991bg. Instead of SN\,1991bg, we plot SN\,2005bl \citep{Hachinger09} which is a similar SN. Right: Nebular phase spectra (black) and models  (blue) 
 of the same SNe.}
\label{fig:sublum}
\end{figure*}

Having examined two transitional SNe Ia in detail we now look at the general properties of transitional 
and sub-luminous SNe Ia. 
It has been suggested that sub-luminous SNe Ia could come from multiple progenitor scenarios, and could be a totally different class of explosion when compared to normal and transitional SNe Ia
\citep[see, \eg][]{WDWD,Mazzali08, Sim10,Blondin17,Jiang17}. Furthermore, it has been claimed by some that SNe Ia with \Dm>1.4\,mag can can only be produced by sub Ch-mass explosions, and not delayed detonation explosions \citep{Blondin17}.
Here we address the evidence for this at the faint end of the luminosity width relation,
and examine how the progenitors of transitional SNe Ia link to those of normal and sub-luminous SNe Ia. 
To do this we examine six SNe Ia (from most to least luminous: SNe~2004eo, 2003hv, 2011iv, 2007on, 1986G, 1991bg) which have extensive and accurate datasets and that have been studied in detail through spectral modelling. These SNe
  vary in $s_{BV}$ between 0.81 and 0.38.  
Table \ref{table:transi} shows the basic properties of each SN, as well as their 
suggested explosion scenario, and Figure \ref{fig:sublum} shows the maximum light and nebular phase spectra and models of these SNe.

\begin{table*}
 \centering
 \caption{The main properties of six SNe Ia including  normal (SN\,2004eo), transitional (SN\,1986G, SN\,2003hv, SN\,2007on, SN\,2011iv) and sub-luminous (SN\,1991bg) SNe Ia. All of the SN in the table 
 have been spectroscopically moddeled, and the \Nifs\ mass was derived from spectral modelling. The SN are ordered in luminosity.
  Table references: (1) \citet{Mazzali08}, (2) \citet{Mazzali11}, (3) This paper, (4) \citet{Ashall16b}, (5) \citet{Mazzali12}.
 }
 \begin{minipage}{250mm}
  \begin{tabular}{cccccccccccc}
  \hline
SN&$s_{BV}$&\Dm&Bol $L_{max}$&\Nifs&Explosion &\Mej&Host type& Observational Properties&Ref.\\ 
&&mag&$10^{43}$erg&$\Msun$&&$\Msun$&&&\\
  \hline
2004eo    &0.81&1.46 &1.15& 0.43&DDT\footnote{Delayed detonation.}&$\sim1.38$&SB&Normal SNe Ia&1\\
2003hv    &  0.75& 1.61&1.00&0.2-0.4\footnote{Dependent on rise time assumed.}&sub Ch det./DDT$^{c}$&$\sim1.2/1.38$&E/S0&High [FeII] to [FeIII] ratio&2\\
2011iv     &0.65&1.77&0.87&0.31&DDT$^a$\footnote{Off-centre detonation.}\footnote{\Nifs\ made in deflagration phase.}&$\sim1.38$&E&Luminous for LC shape&3\\
2007on   &	0.58&1.96&0.54&0.25&Collision/DDT$^a$$^{b}$&1.42/1.38&E&Double peak nebular lines&3\\
1986G     &	0.54&1.81	 &0.4&0.14&DDT$^a$\footnote{No \Nifs\ made in deflagration phase.}&$\sim1.38$&S0-pec&Intermediate Ti II absorption&4\\
1991bg   &	0.38&1.98	&0.26&0.1&Merger/DDT$^a$&$\sim1.2-1.38$&E&Rapid ionisation change\footnote{In the nebular phase.}&5\\
\hline
\end{tabular}
 \end{minipage}
\label{table:transi}
\end{table*}


SN\,2004eo is on the faint end of the normal SNe Ia population; it was  consistent with 
a Ch-mass delayed detonation explosion that produced 0.43\,$\Msun$ of \Nifs\ \citep{Mazzali08}.
Moving down in luminosity, SN\,2003hv  had an early time evolution between that of SNe~2004eo and 
 2011iv.  Optical nebular phase modelling found that SN\,2003hv's high [FeIII] to [FeII] ratio
 was caused by a lower $\rho_{c}$, and less recombination than a standard SNe Ia \citep{Mazzali11}.
This is in line with density profiles of sub Ch-mass models that have an ejecta mass of  $\sim1.2\,\Msun$.
However, mid-infrared observations indicate that SN\,2003hv shows signs of asymmetry  \citep{Gerardy07}. 
 \citet{Motohara06} show that, at +394\,d relative to $B$-band maximum, 
 the 1.257$\mu$m and 1.644$\mu$m [\FeII] features had a pot-bellied profile shape \citep[see][Fig.~13 for details]{Stritzinger15}. This indicates an 
 off-centre explosion, and that the highest density burning in this SN took place quite far away 
 from the centre of  the progenitor star, implying that SN\,2003hv could have been a low central
 density, delayed detonation Ch-mass explosion.
  
SN\,2011iv had an $s_{BV}$ of 0.65 and produced fits which are in line with a
 delayed detonation explosion, which had a higher $\rho_{c}$ and a
  lower $\rho_{tr}$ than a standard SNe Ia. 
 It is possible that SN\,2011iv had an an off-centre ignition \citep{Mazzali17}. 
The next SN in the sequence is SN\,2007on. 
This SN deviates from the standard explosion model, as is seen from 
its double peaked nebular lines. These double peaks possibly come from two separate nebular components \citep{Mazzali17},
indicating the SN could have come from a Ch-mass off-centre ignition delayed 
detonation  explosion with a low $\rho_{c}$, or the collision of two roughly equal  mass WDs. 
 SN\,1986G has been studied in detail by \citet{Ashall16b}. This SN is of particular interest as it is one of the few
 SN Ia that has an  early-time, intermediate \TiII\ absorption feature at 4500\AA. It also has a \Nifs\ mass
 between SN\,2007on and SN\,1991bg. SN\,1986G was compatible with a low energy (high $\rho_{c}$)
 delayed detonation Ch-mass explosion.
  
 Finally we turn to SN\,1991bg, the least luminous object in our sample. It produced $\sim$0.1\,$\Msun$\ of \Nifs,
 and was theorised to be a merger of two WDs \citep{Mazzali12}, a sub Ch-mass explosion \citep{Sim10,Blondin17}, 
 or a Ch-mass delayed detonation explosion  \citep{Hoflich02}.
 \citet{Blondin17} argue that  classical delayed detonation models can only follow the luminosity width relation 
 at values of \Dm<1.38\,mag. Further down the relation their theoretical models take a sharp orthogonal-like 
 turn which produces an `anti-relation'. A similar anti-correlation was found by \citet{Pinto00a}, but 
 was subsequently corrected in \citet{Pinto00b}.
Overall \citet{Hoflich17} demonstrated that the sub Ch-mass explosions, or merger models do not correctly reproduce the whole evolution of sub-luminous SNe,
 while delayed detonation models can.
  \citet{Hoflich02} argue that SN\,1999by (a 1991bg-like SN)
 can be explained by a delayed detonation scenario. 
 However, \citet{Blondin17B} claim the opposite result, and that SN\,1999by can only
  be produced with a sub Ch-mass explosion.
It should be noted that the models from \citet{Blondin17B} are too red with respect to the observations. Furthermore,
  their  models predict a \Dm\ of  1.64\, mag, which is considerably different from the measured value of \Dm=1.90\,mag \citep{Garnavich04}.  
On the other hand,  SN\,1991bg models do require a decrease in density in the inner most region 
of the ejecta to explain the rapid ionisation change in the nebular phase \citep{Mazzali12}, and this is not consistent with  delayed detonation model predictions.

 \subsection{$s_{BV}$ vs. $L_{max}$ relation}
 It is apparent that 
 most transitional and sub-luminous SNe Ia can be modelled by Ch-mass delayed detonation explosions,
but this does not rule out other progenitor or explosion scenarios.
 Regardless of the nature of these explosions, 
 the observational properties of SNe Ia at the faint end of the luminosity width relation are diverse. 
 
 The $s_{BV}$ parameter has allowed the area of fast declining SNe Ia to be reassessed.
Recently, it has been suggested that some sub-luminous objects come from a distinct
 population compared to normal SNe Ia.
 By examining $s_{BV}$ vs. bolometric maximum light luminosity  ($L_{max}$) \citet{Dhawan17} determined that 6 out of 24 SNe Ia came from a different class of explosions. 
 These SNe Ia were the least luminous in the sample, had no near-infrared secondary maximum and had a near-infrared primary maximum after 
 the time of $B$-band maximum. Whereas normal SNe Ia have two near-infrared maxima, one before $B$-band maximum and one
 after $B$-band maximum. Typically, observations show that as the luminosity 
 of  SNe Ia decreases the  time of secondary maximum gets closer to the time of $B$-band maximum, except for the least luminous
 SNe Ia which show one near-infrared maximum after the time of $B$-band maximum \citep{Phillips12}. 

We examine the evidence for two populations of sub-luminous SNe Ia with our models and results. 
To do this we take 24 SNe Ia from the \textit{CSP} and compute their bolometric 
light curves to obtain $L_{max}$. 
The bolometric light curves are constructed from the CSP data using all available filters: $ugriBVYJH$. 
Since the photometry are not all coincident, we must first fit interpolating functions to the observed data.
In order to capture the uncertainty in the interpolation process, we use Gaussian Processes (GP) \citep{Rasmussen06} with a Matern kernel. The mean function is taken to be the best-fit from SNooPy \citep{Burns11,Burns14},
with a linear extrapolation at late times if data exist beyond the extent of the template. This allows for a 
robust fit of the light-curves, but also captures the diversity in the light-curves shapes from SN to SN.

With the measured and interpolated fluxes, we multiply the \citet{Hsiao07} spectral energy distribution (SED) with 
a basis spline and tune the spline coefficients until synthetic photometry from the SED matches the observed 
colours. Using the methods from \citet{Burns14}, we estimate the extinction due to both Milky-Way and host
 galaxy dust. The warped SED is then de-reddened using the \citet{Fitzpatrick99} reddening law. Finally,
 the warped, de-reddened SED is integrated from $\lambda_u$ to $\lambda_H$ to get the bolometric flux. 
The bolometric flux is converted into a bolometric luminosity assuming a Hubble distance with $H_0 = 70 km\,s^{-1}Mpc^{-1}$, or using an independent distance estimate for nearby SNe (see Table \ref{table:Lbol}). 

To determine uncertainties in $L_{bol}$, we take a Monte carlo approach. We draw 100 random realisations 
from the Gaussian Process interpolators and 100 random extinctions consistent with the observed colours and
compute 100 values of the bolometric light curve. This simulates the propagation of correlated errors from
 both the interpolating in time (GPs) and in wavelength (de-reddening) to the bolometric flux.
We find the errors due to extinction and distance to be the most important. A spline function is  fitted 
to the  bolometric light curves to obtain the bolometric maximum, $L_{max}$, for each MC iteration.
The median and standard deviation of the bolometric maximum 
 from the 100 MC iterations is used as the
final bolometric maximum and its associated error. Table \ref{table:Lbol} presents the properties
and values of the 24 SNe analysed.

\begin{figure*}
\centering
\includegraphics[scale=0.5]{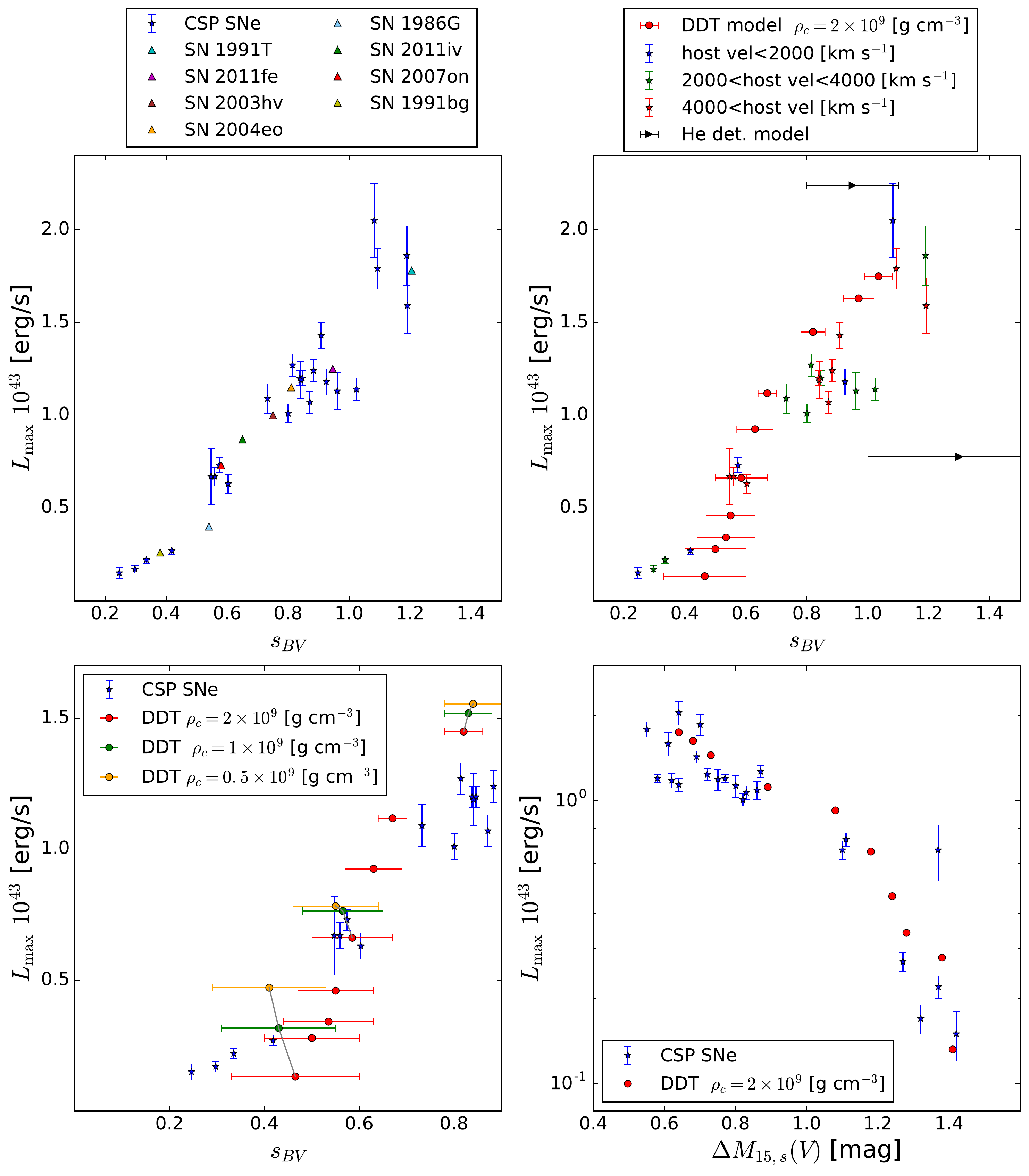}
\caption{{\it Top left: } $s_{BV}$ vs. $L_{max}$   relation for the 24 CSP SNe Ia (blue stars), and 8 SNe modelled through abundance tomography (triangles). 
{\it Top right: } $s_{BV}$ vs. $L_{max}$  relation for the same 24 CSP SNe (stars) and explosion models.
  The SNe are separated in colour depending on the velocity of their host galaxy.
  The bins correspond to SNe which were
 in galaxies with velocities less than 2000\,\kms (red),  with velocities between 2000-4000\,\kms (green), 
 and with velocities greater than 4000\,\kms(orange). 
  The red circles are delayed detonation  models with the same initial properties but varying $\rho_{tr}$, and 
  the black triangles and He detonation models. 
  {\it Bottom left: } The same as the top right panel but zoomed on the sub-luminous section, 
  delayed detonation  models with different $\rho_{c}$ 1$\times10^{9}$\,gcm$^{-3}$ (green circle) 
  and 0.5$\times10^{9}$\,gcm$^{-3}$(orange circle) are also plotted. The grey solid line connects models which have the same $\rho_{tr}$.
 {\it Bottom right: } $\Delta M_{15,s}(V)$ vs. $L_{max}$ for comparison, $\Delta M_{15,s}(V)$ is the preferred 
 measurement of the light curve in \citet{Hoflich17}. Note that  the errors on the observations do not include errors on the distances, see Table \ref{table:Lbol}. The errors in the theoretical $s_{BV}$ measurment
are caused by a  jitter in $B$ and $V$ due to the spatial discretization
of the models and uncertainties in the atomic models, in particular
during the phase of the Lira-relation, as discussed in \citet{Hoflich17}.}
\label{fig:DDT}
\end{figure*}

The top left panel in  Figure \ref{fig:DDT} shows  $s_{BV}$ vs  $L_{max}$
for the 24 CSP SNe Ia we have analysed, and 
seven SNe Ia which were modelled using the abundance tomography approach. 
From the CSP data alone we find no objects in the area of parameter  space (i.e., $0.45<s_{BV}<0.6$) 
between the least luminous and 'normal' SNe Ia populations. 
However from the abundance stratification models we find that 
 SN\,1986G sits in the area of parameter space between the two populations and acts as a link between normal and sub-luminous SNe Ia.
 This is where one would expect SN\,1986G to be located, as it is the only SNe\,Ia which had an intermediate \TiII\ absorption feature.
 Furthermore, SN\,1986G had a secondary near-infrared maximum 10$\pm$5\,d\footnote{Generous errors are given due to the age of the photometry.}
after  $B$-band maximum  \citep{Frogel87}. This is significantly shorter than the time of secondary near-infrared maximum
 of other SNe Ia. For example the second near-infrared maximum of 
 SN~2007on occurred 20 days after $B$-band maximum.  
 However, due to its large host galaxy extinction, the position of  SN\,1986G on this plot is very uncertain.  
It is apparent from Figure \ref{fig:DDT} that these connecting 1986G-like SNe are rare.  
Another SN which could possibly be a connection between these two populations is SN\,2007ba.
Spectroscopically, it was classified as a 1991bg-like SN, it had strong \TiII\ absorption at maximum 
light, and a near-infrared secondary maximum, but the time of $Y$ band maximum  was after the time of $B$-band 
maximum. This is not the case for the other normal SNe Ia. It should also be noted that the host galaxy 
extinction calculated for SN\,2007ba in \citet{Burns14} is 
too large and incompatible with its spectral shape. In fact the spectra of SN\,2007ba indicate 
that the SN is consistent with almost no host galaxy extinction. 
The likely reason for this incorrect value of extinction is because it was assumed that there was a single intrinsic 
colour-$s_{BV}$ relation for SNe Ia. However,  according to theoretical models for fast declining SNe Ia 
there should be a variation in colour, caused by different $\rho_{c}$, for the same $s_{BV}$. 

To examine the $s_{BV}$ vs. $L_{max}$ relation further we turn our attention to the delayed detonation models of \citet{Hoflich17} (see Table \ref{table:models}). 
The  model light curves are based on time-dependent non-LTE simulations, as described in \citet{Hoeflich95,Hoflich02,Hoflich03a,Hoflich03b} and references there in. The models were upgraded by atomic models subsequently.
They are Ch-mass explosions of carbon-oxygen WDs. 
After an initial deflagration phase, a detonation is triggered when the flame front drops to a $\rho_{tr}$.
For a normal bright SNe Ia, to first order, the $\rho_{tr}$ determines the amount of \Nifs\ synthesised in the ejecta, 
as there is more effective burning during the detonation. Hence,  $\rho_{tr}$ is the main driver in the luminosity width relation \citep{Hoflich02}.
However, some second order parameters that can affect the evolution of the explosion are the primordial metalicity of the progenitor star, the size of the accompanying main sequence star, and the $\rho_{c}$  of the WD. Here we will just focus on $\rho_{tr}$ and $\rho_{c}$.

Comparing  the models of \citet{Hoflich17} to the CSP data we find a very similar trend.
In the top right panel of Figure \ref{fig:DDT} we present a  suite of models with the same central 
 density ($2\times10^{9}$g\,cm$^{-3}$), solar metalicity and main sequence mass of 5\,$\Msun$, but varying 
 $\rho_{tr}$.   To first order it is clear that varying  $\rho_{tr}$ produces the 
 same trend as the observations. 
 The sudden drop in luminosity at $s_{BV}$ =0.4--0.6 corresponds to the \Dm `cliff'. 
 The  \Dm `cliff' is seen as a rapid drop in brightness at \Dm$>$1.5\,mag, it is caused by 
 the transition from doubly to singly ionised iron group elements  which allows SNe to enter a regime of quickly dropping opacity, 
(for more discussion see \citet{Hoflich17}).
In these models the least luminous SNe show only one near-infrared maximum (which peaks after $B$-band maximum)
whereas the more luminous models show two near-infrared maxima (one peaks before $B$-band maximum and one after). 
This is because the secondary peak is related to the phase of the
shrinking photospheric radius \citep{Hoeflich95}. Therefore, the secondary peak is later as brightness increases.
 In the sub-luminous regime the first and second peak appear to merge and show one very broad peak, which is always 
later than the first peak in the more luminous models.
 For comparison, we also plot two He detonation models from \citet{Hoflich17} which appear not to 
  match the observations.
We emphasise that the models presented here were not tuned to fit the data, and all models were chosen from previously published work. 
 
 For normal SNe Ia,  10-20\% of \Nifs\ can be made in the deflagration phase, where some high-density deflagration burning
 is required for the pre-expansion to agree with observations. These densities are high enough  to 
 reach NSE, and when the densities exceed \ab$10^9$\,g\,cm$^{-3}$ 
it can cause a shift in the NSE away from \Nifs\ towards stable NSE elements, due to the increased electron capture rate \citep{Hoflich17}. 
Hence, the amount of \Nifs\ produced during the deflagration phase is dependent on  $\rho_{c}$. 
 For sub-luminous SNe Ia deflagration burning dominates and most of the \Nifs\ is produced during this phase.
 Therefore, the $\rho_{c}$  of sub-luminous SNe Ia has a large affect on its luminosity. In fact, a change 
 of $\rho_{c}$=$1\times 10^9$\,g\,cm$^{-3}$ to $2\times 10^9$\,g\,cm$^{-3}$ can change the  \Nifs\ mass by a factor of 2, which could explain some of the difference between SN~1986G and SN~2007on. 
Generally, a SN Ia with a high  $\rho_{c}$ will not have \Nifs\ in the inner most layers, it will be located further out in the ejecta \citep{Hoflich17}, as
is seen in the abundance stratification models of SN\,2011iv and SN\,1986G (see Figure \ref{fig:tomo}).
The bottom left panel in Figure \ref{fig:DDT} is similar to the 
 top right panel but with delayed detonation models which have different $\rho_{c}$. 
As discussed above, changing the $\rho_{c}$ has the biggest affect on the least luminous models.  
 
It is evident from the explosion and spectral models that SNe Ia can be produced through the delayed detonation scenario 
 for almost the whole range of observations. However, in nature we do not see a smooth distribution
 of SNe Ia properties. There is a bimodal distribution in $L_{max}$, with a lack of SNe Ia connecting sub-luminous and 
  normal SNe Ia. 
  The increased number of objects at the faint end of the distribution could be  due to 
  multiple progenitor scenarios (delayed detonation explosions, small amplitude PDDs, mergers and sub Ch-mass explosion).
  There is clearly a lack of SNe in the region between normal and sub-luminous SNe Ia, and it is likely to be intrinsic
 as only a handful of these objects have ever been discovered and followed.\footnote{There could be 
   an observer bias where only 1991bg-like SNe have been followed as these were deemed more interesting.} 
This region 
corresponds to a $\rho_{tr}=\sim12-14\times10^{6}$g\,cm$^{-3}$. 
 To truly assess the rate of these objects
 a large volume limited sample is needed.

\section{Conclusion}
We have modelled two transitional SNe Ia (SNe\,2007on and 2011iv) using the abundance stratification technique. 
SNe\,2007on and 2011iv both exploded in NGC\,1404 at a distance modulus of 31.2$\pm$0.2\,mag.
As well as being in the same galaxy, these SNe sit in an important area of parameter space. 
Understanding how they exploded will let us determine the connection between normal and sub-luminous SNe Ia. 
SN\,2011iv was a low energy (high $\rho_{c}$) Ch-mass explosion. 
It synthesised $0.31\,\Msun$ \Nifs, and  had a \KE\ of $1.2\pm0.3\times10^{51}$\,erg. 
We favour a low energy Ch-mass explosion model (W7e0.7) as it produces good spectral fits, and
provides results which are roughly consistent  between the \KE\ of the input density profile and the \KE\ calculated from the integrated abundances. 
W7e0.7 is also the best fit density profile best suited to SN\,1986G (another transitional SN Ia; \citet{Ashall16b}). 
Our results are consistent with SN 2011iv being a delayed detonation Ch-mass explosion with
a low $\rho_{tr}$.

SN\,2007on was less luminous than SN\,2011iv but neither of them showed the \TiII\ absorption at maximum light, which is 
indicative of a sub-luminous SN Ia.  The late-time spectra of 
SN\,2007on show two separate components, see \citet{Mazzali17}. Therefore we do not produce
a 1-zone nebular model and `complete' the abundance stratification experiment.  
\citet{Mazzali17} demonstrate that these components could come from either 
the  collision of two WDs, or a very off-centre ignition in the delayed detonation scenario.
However, at early times the two components are significantly below our photosphere
so the early time results in this work are still valid. 
Photospheric phase models were produced for SN\,2007on
using  the W7, W7e0.7 and W7e2 density profiles. The W7 density profile
produces marginally better fits, as the earlier fits from the W7e0.7 model were too hot, 
and those from W7e2 had too many high velocity lines.

\vspace{10mm}
If SN\,2007on was a delayed detonation Ch-mass explosion, the fact that the W7 model produces the best fits indicates that 
it had a lower  $\rho_{c}$ than SN\,2011iv. 
 This agrees with the results of \citet{Gall17}.
However, the  lack of stable NSE elements (see Section 5.1 and \citet{Mazzali17})  in the ejecta of SN~2007on may 
 favour the  collision of two \ab0.7\,$\Msun$ WDs.
Overall, as shown in Section 5.3, our spectral modelling results indicate that SNe~1986G, 2007on 
and 2011iv were consistent with being produced from a delayed detonation  explosion where SNe~1986G and 2007on
 had a similar $\rho_{tr}$ 
but different $\rho_{c}$, and SNe~2011iv and 1986G had a similar $\rho_{c}$ 
but different $\rho_{tr}$.

To examine the area of transitional and sub-luminous SNe Ia further,
we turned to previously published spectral models.
 SN\,2004eo (\Dm=1.46\,mag) was proposed to be a standard delayed detonation Ch-mass explosion \citep{Mazzali08}.
 SN\,2003hv (\Dm=1.61\,mag) is compatible with either showed signs of  a sub Ch-mass detonation or a 
 delayed detonation Ch-mass explosion,
 and  SN\,1991bg (\Dm=1.98\,mag) could have been a sub Ch-mass detonation, a merger of two WDs,
 or a Ch-mass delayed detonation explosion.
 Either way, the diversity of SNe Ia at the faint end of the luminosity width  relation is large, and 
 there are probably different explosion mechanisms, and possibly different progenitor channels, at work. 

The introduction of the $s_{BV}$ parameter has allowed fast declining SNe to become more standardisable
 (Burns et al, in prep). 
It also provides a new parameter with which we can reaccess transitional SNe Ia.
To do this we computed the peak bolometric maximum ($L_{max}$) for 24 SNe Ia, which were
observed during the CSP.   
A large drop in the  $s_{BV}$ vs $L_{max}$ relationship was found at $s_{BV}=\sim0.5$.
Some have interpreted this as being an indication that there are two distinct populations of SNe Ia. 
However, we show that SN\,1986G sits in this `gap', and links sub-luminous
and normal SNe Ia populations. 

To explain the observational trend we turn to the published  models of \citet{Hoflich17}.
These models are delayed detonation Ch-mass explosions. 
To first order the behavior of SNe Ia, including that of the sub-luminous SNe Ia,
 can be explained by varying the $\rho_{tr}$ in the explosion. The largest second order affect
 for sub-luminous SNe Ia is the $\rho_{c}$ of the WD. 
 For normal SNe Ia, changes in $\rho_{c}$ are insignificant. For 
 example a factor of 2 change will produce a change in \Nifs\ mass of  10-20\%.  
 For  sub-luminous SNe Ia $\rho_{c}$ has a large affect, due to most of the \Nifs\ being produced in the deflagration phase. 
 A factor of 2 change in $\rho_{c}$ for the least luminous SNe will produce a factor of 2 change in \Nifs\ mass. 
However, we note that just because the models fit almost all of the data does not mean that they are the only 
progenitor scenario we observe. There are  hints, with supernovae such as 
SNe\,2007on and 1991bg, of different progenitor and explosion channels, and
it is likely that most progenitor scenarios do exist. However, currently the strongest 
evidence seems to suggest that the bulk of SNe Ia come from a one parameter family.

In conclusion, we find evidence that SNe\,2007on and 2011iv 
both had ejecta masses similar to the Ch-mass and experienced a  delayed detonation explosion. Although we note the possibility that SN~2007on could have originated from the collision of two similar mass WDs \citep[\eg][]{Mazzali17}.
Our overview of the faint end of the luminosity width relation has shown that there could be a smooth 
continuum between normal and sub-luminous SNe~Ia, which to first order can be explained by varying $\rho_{tr}$ in the delayed detonation explosion scenario.
Looking forward to the future, 
we now have data sets where physical second order effects can be examined. 
We need to explore the edges of the parameter space, with models (both abundance stratification and explosion) that produce less \Nifs,
and with models which explore alternative explosion scenarios more robustly. 
However, it appears that even at the faint end of SNe Ia distribution most explosions come from 
Ch-mass objects.

\section*{Acknowledgments}
C. A. would like to acknowledge support by the NSF research grant  AST-1613472.
M. D. S. acknowledges support by a research grant (13261) from VILLUM FONDEN. 
The work presented in this paper has been supported in part by 
NSF awards AST-1613472 (PI: E.~Y. Hsiao), P. Hoeflich AST-1715133 (PI: P. Hoeflich),
 and AST-1613426 (PI: M.~M. Phillips).
We  also acknowledge  the Florida Space Grant Consortium.
C.G. acknowledges support by the Carlsberg Foundation.
The UCSC team is supported in part by NSF grant AST-1518052, the
Gordon \& Betty Moore Foundation, and from fellowships from the Alfred
P.\ Sloan Foundation and the David and Lucile Packard Foundation to
R.J.F.
Support for program \#GO--12592 was provided by NASA through a grant from the Space Telescope Science Institute, which is operated by the Association of Universities for Research in Astronomy, Inc., under NASA contract NAS 5--26555.

\appendix

\section{Distance to SN~2007on and SN~2011iv} \label{app:dista}
\begin{figure*}
\centering
\includegraphics[scale=0.5]{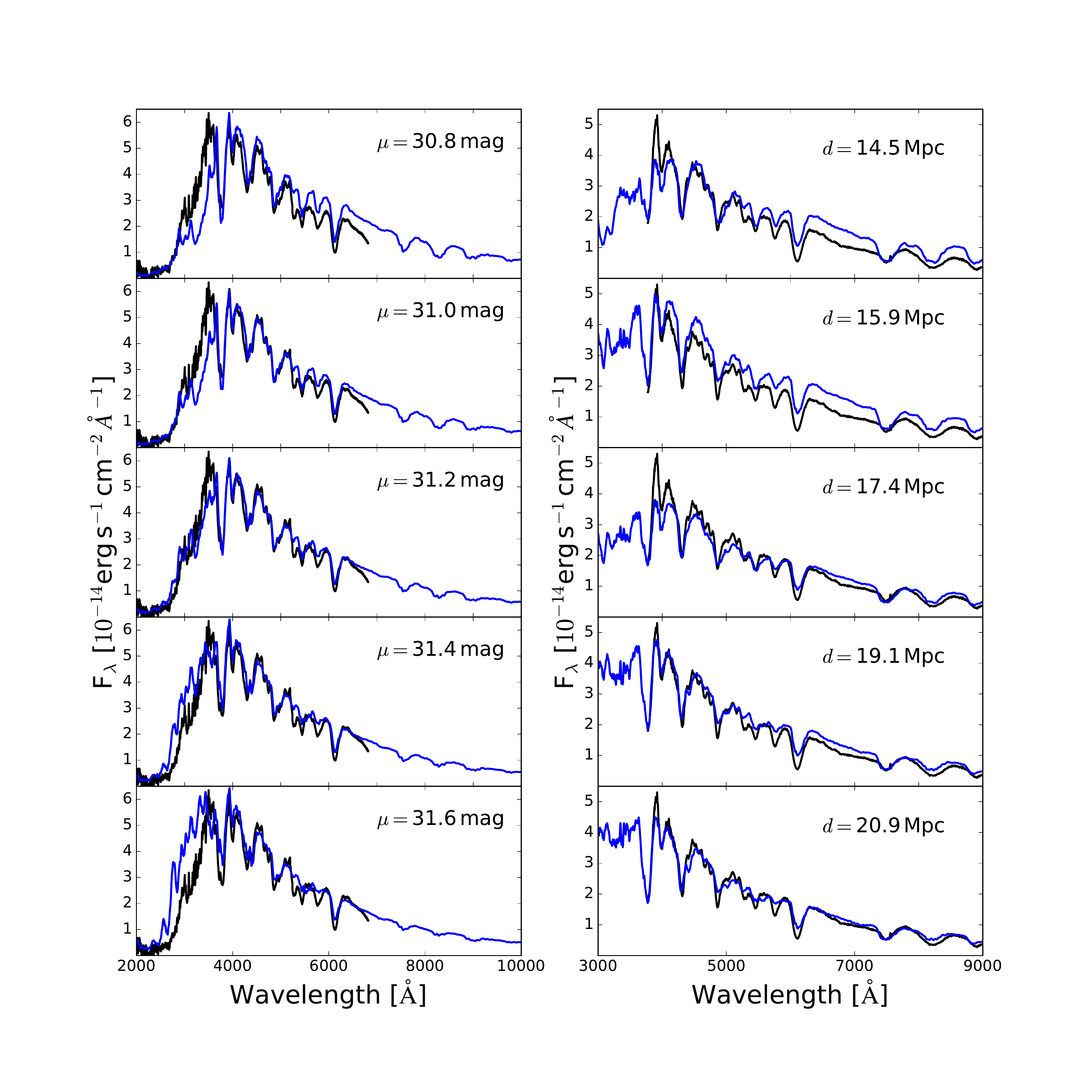}
\caption{Spectra (black) and models (blue) of SN\,2011iv (left) and SN\,2007on (right) at --4.4\,d and --4.0\,d
relative to $B$-band maximum. The models have been made for a range of distance moduli (i.e., 30.8 $-$ 31.6\,mag).}
\label{fig:distance}
\end{figure*}

The distance to NGC~1404 is very uncertain and could be anywhere between 13.6 and 27.9\,Mpc.
As SNe\,2007on and 2011iv both exploded in this galaxy we can use spectral modelling 
to quantify this distance. Figure \ref{fig:distance} contains spectral models of SN\,2011iv and SN\,2007on
at $-$5.5\,d and $-$4.0\,d relative to $B$-band maximum, produced for five different distance moduli. 
Here, as in the rest of this paper, we assume no host galaxy extinction. For a discussion on this see \citet{Gall17}.

The models at the larger distance ($\mu$=31.6\,mag or d=20.9\,Mpc) do not produce good fits 
and are too hot. For example the  model of SN~2011iv has a very strong \SiIII\ \lam4553 feature and 
too much flux in the near-UV. If the flux in the near-UV was reduced by adding metals to the outer layers of the ejecta,
it would reprocess this flux into the optical and further increase the temperature and ionisation state, and would therefore worsen the fit. 
A poor fit can be seen in the SN~2007on model where the \SiIII\ \lam4553 feature  is too strong,
and there is a weak Si ($\sim$5970 and 6355\AA) ratio. Furthermore, the feature at 4900\AA\ at this epoch is dominated 
by \SiII\ \lam\lam5041 5056, but at this large distance the ionization state of Si is too high and there is not enough \SiII\
absorption. Hence, it is apparent that the larger distance requires a luminosity (and therefore temperature)
that is inconsistent with the observations.

 The models at the shortest distance ($\mu$=30.8\,mag or d=14.5\,Mpc) also produce poor fits. 
 For SN~2011iv there is too much absorption in the UV. It also does not contain enough  \FeIII\ \lam5156 absorption in the
 red side of the 4900\AA\ feature. Due to the lower distance and therefore luminosity
 and temperature, \FeII\ \lam\lam5018, 5169 dominates this feature.

Therefore, a distance of  $\mu$=31.6\,mag produces models which require a luminosity which is too high to match the observations,
and  a distance of  $\mu$=30.8\,mag produces models which are require a luminosity which is too low to match the observations.
Hence, the optimum solution is somewhere in-between. The models at $\mu=$31.2\,mag produce the best fits.
For SN~2011iv the UV flux level is good, whilst the fit in the optical part of the spectrum has improved. 
For SN~2007on the line shapes, depths, equivalent widths and ratios are all  remarkably good.
Therefore, we adopt  a distance modulus of $\mu=$31.0$\pm$0.2\,mag. More evidence of this distance 
is that it places SN~2007on and SN~2011iv on the correct part of the luminosity width  relation.

 \section{Log of data}
\begin{table}
 \centering
 \caption{The spectra of SN 2011iv.}
 \begin{minipage}{250mm}
  \begin{tabular}{cccccc}
  \hline
   Date&JD&Phase (days)\footnote{Relative to $B$-band maximum in observer frame.}&telescope\\
   \hline
4/12/2011&2455899.5&$-$6.6&du Pont&B\&C\\
1/12/2011&2455901.0&$-$5.1&Swift&UVOT\\
5/12/2011&2455900.6&$-$5.5&du Pont&B\&C\\
1/12/2011&2455902.1&$-$4.0&Swift&UVOT\\
1/12/2011&2455903.1&$-$3.0&Swift&UVOT\\
1/12/2011&2455905.1&$-$1.0&Swift&UVOT\\
11/12/2011&2455906.5&+0.4&HST&STIS\\
1/12/2011&2455907.2&+1.1&Swift&UVOT\\
11/12/2011&2455910.5&+4.4&HST&STIS\\
11/12/2011&2455915.5&+9.4&HST&STIS\\
11/12/2011&2455919.5&+13.4&HST&STIS\\
22/1/1987&2456166.8&+260.7&NTT&EFOSC\\
\hline
\end{tabular}
\label{table:data11iv}
\end{minipage}
\end{table}

\begin{table}
 \centering
 \caption{The spectra of SN 2007on.}
  \begin{minipage}{250mm}
  \begin{tabular}{cccccc}
  \hline
   Date&JD&Phase (days)\footnote{Relative to $B$-band maximum in observer frame.}&telescope\\
   \hline
11/11/2007&2454415.8&$-$4.0&du Pont&WFCCD\\
14/11/2007&2454418.8&$-$1.0&du Pont&WFCCD\\
17/11/2007&2454421.7&+1.9&Baade&IMACS\\
19/11/2007&24554423.7&+3.9&NTT&EMMI\\
25/11/2007&24554429.7&+9.9&Clay&MagE\\
\hline
\end{tabular}
\label{table:data07on}
 \end{minipage}
\end{table} 

 \section{SNe Ia bolometric maximum data and models.}
 
\begin{table*}
 \centering
 \caption{The properties of the CSP SNe IA used in Figure \ref{fig:DDT}. The distance used to calculate the bolometric luminosity is indicated by *. The different distance methods are TF=Tully Fisher, SBF=Surface Brightness Fluctuation. }
  \begin{minipage}{250mm}
  \begin{tabular}{ccccccccccc}
  \hline
   SN&$L_{max}$&z&Host velocity&LD ($\mu$) \footnote{Luminosity distance corrected to the reference frame defined by the 3K microwave background radiation.}&\Dm\footnote{Obtained from  \textsc{SNooPY}.}&$\Delta m_{15,s}(V)$\footnote{Obtained from \citet{Hoflich17}.}&$s_{BV}$\footnote{Obtained from \citet{Burns14}.}&other $\mu$ (method)&host galaxy\\
   &10$^{43}$\,ergs$^{-1}$&&\kms&&mag&&&&\\
   \hline
2005am&1.09$\pm$0.08&0.0079&2368&32.94*& 1.49&0.86&0.73&32.64$\pm$0.80 (TF)&NGC 2811\\
2005el&1.20$\pm$0.04&0.0149&4470&34.05*&1.36&0.58&0.84&$\cdots$&NGC 1819\\
    2005hc&1.59$\pm$0.15&0.0459&13772&36.5*&0.83&0.61&1.19&$\cdots$&MCG+00-06-003\\
   2005iq&1.07$\pm$0.06&0.0340&10206&35.8*& 1.25&0.83&0.87&$\cdots$&eso538G013\\
  2005ke&0.27$\pm$0.02&0.0049&1463&31.43&1.69&1.27&0.42&31.89$\pm$0.84 (TF)*&NGC 1371\\ 
   2006D&1.27$\pm$0.06&0.0085&2556&33.10*&1.36&0.87&0.81&$\cdots$&MCG -01-33-34\\
   2006bh&1.01$\pm$0.05&0.0108&3253&33.28*& 1.41&0.82&0.80&33.22$\pm$0.1 (TF)&NGC 7329\\
   2006et&1.79$\pm$0.11&0.0226&6787&34.87*&0.84&0.55&1.09&$\cdots$&NGC 0232\\
   2006gt&0.67$\pm$0.05&0.0448&13422&36.43*&1.64&1.10&0.56&$\cdots$&\footnote{2MASXJ00561810-0137327.}\\
   2006mr&0.15$\pm$0.03&0.0059&1760&31.87&1.93&1.42&0.25&31.25$\pm$0.51(TF,SBF)*&NGC 1316\\
   2007N&0.17$\pm$0.02&0.0129&3861&33.91*&2.12&1.32&0.30&34.06$\pm$0.18 (TF)&MGC0133012\\
  2009F&0.22$\pm$0.02&0.0130&3884&33.74*&1.97&1.37&0.33&$\cdots$&NGC 1725\\
  2007af&1.18$\pm$0.07&0.0055&1638&32.16& 1.19&0.62&0.93&31.75$\pm$0.05 (Cephid)*&NGC 5584\\
  2007ba&0.67$\pm$0.02&0.0385&11546&36.18*&1.67&1.37\footnote{This value is uncertain due to poor temporal coverage.}&0.55&$\cdots$&UGC 09798\\
  2007bd&1.24$\pm$0.06&0.0309&9266&35.73*&1.24&0.72&0.88&$\cdots$&UGC 04455\\
   2007hj&0.63$\pm$0.05&0.0141&4231&33.73*&1.63&1.02&0.60&$\cdots$&NGC 7461\\
   2007le&1.14$\pm$0.06&0.0067&2015&31.88&0.93&0.63&1.02&$\cdots$&NGC 7721\\
   2007on&0.73$\pm$0.04&0.0065&1947&32.13&1.96&1.11&0.57&31.2 (\footnote{\cite{Gall17}.}+this paper)*&NGC 1404\\
    2009aa&1.43$\pm$0.07&0.0273&8187&35.48*&1.21&0.69&0.91&-&ESO57G020\\
       2008fp&2.05$\pm$0.20&0.0056&1698&32.16&0.76&0.64&1.08&31.82$\pm$0.01\footnote{\cite{Weyant14}.}*&ESO 428-G14\\
    2008hv&1.20$\pm$0.04&0.0125&3762&33.85*&1.31&0.76&0.85&$\cdots$&NGC 2765\\
   2008ia&1.19$\pm$0.10&0.0219&6578&34.96*&1.23&0.75&0.84&$\cdots$&ESO125006\\
   2009Y&1.86$\pm$0.16&0.0094&2804&33.20*&0.97&0.70 &1.19&32.20$\pm$0.27 (TF)&NGC 5728\\
   2009ag&1.13$\pm$0.10&0.0086&2590&33.01*&1.05&0.80$^{c}$&0.96&32.18$\pm$0.3 (TF)&ESO 492-2\\
\hline
\end{tabular}
\label{table:Lbol}
 \end{minipage}
\end{table*}

\begin{table*}
 \centering
 \caption{Delayed detonation models from \citet{Hoflich17}.}
  \begin{minipage}{250mm}
 \begin{tabular}{ccccc}
  \hline
   Model&$log(L_{max})$&$s_{BV}$&$\rho_{tr}$&$\rho_{c}$\\
   \hline & erg~s$^{-1}$&&$10^{7}$\,g~cm$^{-3}$&$10^{9}$\,g~cm$^{-3}$\\
      \hline
8&42.21&0.33--0.60&8&2\\
8r1&42.54&0.31--0.55&8&1\\
8r2&42.70&0.29--0.53&8&0.5\\
10&42.49&0.40--0.60&10&2\\
12&42.57&0.44--0.63&12&2\\
14&42.69&0.47--0.63&14&2\\
16&42.84&0.50--0.67&16&2\\
16r1&42.90&0.48--0.65&16&1\\
16r2&42.91&0.46--0.64&16&0.5\\
18&42.98&0.57--0.69&18&2\\
20&43.06&0.64--0.70&20&2\\
23&43.17&0.78--0.86&23&2\\
23d1&43.20&0.78--0.90&23&0.5\\
23d2&43.19&0.78--0.88&23&1.1\\
25&43.22&0.92--1.02&25&2\\
27&43.25&0.99--1.08&27&2\\
HeD6&42.89&1.00--1.60&n/a&0.01\\
HeD10&43.35&0.80--1.10&n/a&0.03\\
\hline
\end{tabular}
\label{table:models}
 \end{minipage}
\end{table*}

\end{document}